\documentclass[journal]{IEEEtran}
\usepackage{cite}
\usepackage{amsmath,amssymb,amsfonts}
\usepackage{graphicx}
\usepackage{textcomp}
\usepackage{wrapfig}
\usepackage{comment}

\usepackage{multirow}
\usepackage{url}
\usepackage{color}
\usepackage{booktabs} 
\usepackage[caption=false,font=normalsize,labelfont=sf,textfont=sf]{subfig}
\usepackage{stfloats}
\usepackage{verbatim}
\usepackage{algpseudocode}
\usepackage{algorithm}

\def\BibTeX{{\rm B\kern-.05em{\sc i\kern-.025em b}\kern-.08em
    T\kern-.1667em\lower.7ex\hbox{E}\kern-.125emX}}
\definecolor{abstractbg}{rgb}{0.89804,0.94510,0.83137}
\begin{document}
\title{NGGAN: Noise Generation GAN Based on the Practical Measurement Dataset for Narrowband Powerline Communications}
\author{{Ying-Ren~Chien,~\IEEEmembership{Senior~Member,~IEEE}, Po-Heng Chou,~\IEEEmembership{Member,~IEEE}, You-Jie Peng, Chun-Yuan Huang, Hen-Wai Tsao, and~Yu Tsao~\IEEEmembership{Senior Member,~IEEE}}
\thanks{Manuscript received April 9, 2024; revised August 18, 2024; accepted September 6, 2024. This work was supported in part by the National Science and Technology Council (NSTC) of Taiwan under Grants 109-2221-E-197-026, 112-2221-E-197-022, and 113-2926-I-001-502-G, and Academia Sinica, under Grant 235g Postdoctoral Scholar Program \emph{(Corresponding author: Po-Heng Chou)}.}
\thanks{Ying-Ren Chien is with the Adaptive \& Autonomous Communication Lab., Department of Electronic Engineering, National Taipei University of Technology (NTUT), Taipei 10608, Taiwan (e-mail: yrchien@ntut.edu.tw).}
\thanks{Po-Heng Chou and Yu Tsao are with the Research Center for Information Technology Innovation (CITI), Academia Sinica, Taipei, 11529, Taiwan (e-mail: d00942015@ntu.edu.tw;  yu.tsao@citi.sinica.edu.tw).}
\thanks{You-Jie Peng and Hen-Wai Tsao are with the Graduate Institute of Communication Engineering, College of Electrical Engineering and Computer Science (GICE), National Taiwan University (NTU), Taipei, 10617, Taiwan (e-mail: roger851122@gmail.com; tsaohw@ntu.edu.tw).}
\thanks{Chun-Yuan Huang is with the Institute of Communication Engineering (ICE), National Sun Yat-sen University (NSYSU), Kaohsiung, 80424, Taiwan (e-mail: sdff6842@gmail.com).}
}

\markboth{IEEE Transactions on Instrumentation and Measurement,~Vol.~74,~2025}{}

\maketitle

\begin{abstract}
To effectively process impulse noise for narrowband powerline communications (NB-PLCs) transceivers, capturing comprehensive statistics of nonperiodic asynchronous impulsive noise (APIN) is a critical task. However, existing mathematical noise generative models only capture part of the characteristics of noise. In this study, we propose a novel generative adversarial network (GAN) called noise generation GAN (NGGAN) that learns the complicated characteristics of practically measured noise samples for data synthesis. To closely match the statistics of complicated noise over the NB-PLC systems, we measured the NB-PLC noise via the analog coupling and bandpass filtering circuits of a commercial NB-PLC modem to build a realistic dataset. To train NGGAN, we adhere to the following principles: 1) we design the length of input signals that the NGGAN model can fit to facilitate cyclostationary noise generation; 2) the Wasserstein distance is used as a loss function to enhance the similarity between the generated noise and training data; and 3) to measure the similarity performances of GAN-based models based on the mathematical and practically measured datasets, we conduct both quantitative and qualitative analyses. The training datasets include: 1) a piecewise spectral cyclostationary Gaussian model (PSCGM); 2) a frequency-shift (FRESH) filter; and 3) practical measurements from NB-PLC systems. Simulation results demonstrate that the generated noise samples from the proposed NGGAN are highly close to the real noise samples. The principal component analysis (PCA) scatter plots and Fréchet inception distance (FID) analysis have shown that NGGAN outperforms other GAN-based models by generating noise samples with superior fidelity and higher diversity.
\end{abstract}


\section{Introduction}
\label{sec:intro}
\IEEEPARstart{N}{arrowband}
powerline communication (NB-PLC)~\cite{Aderibole2023, Artale2013, Cataliotti2012} is a potential physical layer solution for smart grids, smart homes, and indoor positioning applications~\cite{Antoniali2014}.
However, NB-PLC suffers from noise because it is designed for power delivery rather than signal transmission~\cite{Angrisani2015}.
Measuring complicated noise in NB-PLC is essential for describing the noise model.
A lax and unrealistic noise model may lead to overly optimistic performance concerning detection probability~\cite{Aderibole2023}, data rate~\cite{Artale2013}, and attenuation~\cite{Artale2013, Cataliotti2012}.
Therefore,  it is important to model NB-PLC noise to facilitate the physical design of transceivers~\cite{Omri2020} and network protocols~\cite{Uribe2017} so that robustness against noise is objectively evaluated~\cite{Rouissi2019}.
Channel disturbances in NB-PLC systems were explored in~\cite{Llano2019} and~\cite{ Roopesh2019}. When employing different additive noise models, the corresponding bit error rate (BER) performance results have been observed~\cite{Angrisani2015}. Consequently, if the noise generation model fails to accurately represent most of the noise, the transceiver design may become excessively optimistic and compromise the robustness.

Additive noises in NB-PLC systems (hundreds of Hz to several MHz) are categorized in~\cite{Zimmermann2002} as follows: (1) colored background noise (CBG), (2) narrowband interference (NBI), (3) periodic impulsive noise asynchronous to the mains frequency (PINS), (4) periodic impulsive noise synchronous with the mains (PINAS), and (5) asynchronous impulsive noise (APIN).
The dominant noise in NB-PLC systems is PINS~\cite{Tucci2017}.
However, CBG and NBI should not be disregarded~\cite{Katayama2006}.
Survey work to model NB-PLC noise was investigated in~\cite{Bai2021}.
Most previous studies used curve-fitting to establish mathematical models of NB-PLC noise.
First, the CBG is characterized by one or two dimensions.
In the one-dimensional model of CBG, a zero-mean Gaussian distribution with frequency-dependent variance is used to model the probability density function (PDF).
In the two-dimensional CBG, Rayleigh or Nakagami-m distributions were used to model the PDF in the time domain, and a negative exponential decay form was used to fit the power spectrum density (PSD) in the frequency domain.
Next, the NBI was characterized by a log-normal distribution in the time domain, and the PSD was modeled as a sum of multiple Gaussian-like functions in the frequency domain.
PINS or PINAS can then be modeled using a cyclo-stationary Gaussian process generated from a set of frequency-shift (FRESH) filters~\cite{Elgenedy2016} or a set of parameterized spectral and temporal shaping filters~\cite{Moaveninejad2020}.
\IEEEpubidadjcol
In contrast to PINS and PINAS, the duration and inter-arrival time of APIN are random variables.
Thus, APIN has a high degree of random variability.

However, existing studies can only capture some of the characteristics of additive noise. 
Because non-periodic asynchronous noise exhibits a high degree of random variability in duration and inter-arrival time, to the best of our knowledge, there is no single model that can be used to represent all the types of noise mentioned above. Even if it is possible to combine individual models into a composite model, time synchronization among the sub-models can be a severe problem~\cite{Huang2014}. 
Furthermore, some noise measurements are costly, such as those for multi-phase and large-scale distributed powerline networks. Thus, a data-driven modeling approach is suitable for modeling noise characteristics.
Therefore, it is challenging to accurately capture the trajectories of practical NB-PLC noise using a mathematical model (e.g., the piecewise spectral cyclo-stationary Gaussian model (PSCGM)~\cite{IEEE1901} or FRESH~\cite{Moaveninejad2020}).

Recently, a deep learning (DL) model called a generative adversarial network (GAN)~\cite{Goodfellow2014} has proven to be particularly effective in synthesizing unidentified data from measured data.
The DL-based approach could reduce the cost of some costly measurements that are needed~\cite{Shirmohammadi2021, Khanafer2020}. Examples include electromagnetic interference (EMI)~\cite{Loschi2022}, multi-phase powerline networks~\cite{Bai2021, Elgenedy2018}, long-term noise characterization~\cite{Raponi2022}, large-scale distributed powerline networks~\cite{Sausen2021, Alaya2019}, and cyclic frequency offset issues~\cite{Chien2020}.
When the amount of training data is insufficient, the performance of the DL model is poor, and overfitting easily occurs~\cite{Shirmohammadi2021, Osman2021}.
Therefore, increasing the amount of training data is required to ensure generalizability and avoid overfitting. 
To address this issue, data augmentation using GAN is essential.
In certain applications, such as data processing at end-user devices via mobile edge servers for enhanced quality of service, gathering sufficient training data is challenging owing to privacy concerns and cost constraints~\cite{Pandey2023}.
In addition, the discrepancy between real and generated data can be considered a valuable source of diversity.
To enhance diversity within the training dataset, data augmentation by employing a GAN is also a desirable solution.
Consequently, we emphasized the comparison of the proposed GAN-based model with other GAN-based models. 

To model complex NB-PLC noise, we propose a GAN-based model called the noise-generation GAN (NGGAN), which was inspired by previous works~\cite{Letizia2020, Chien2021}. The proposed NGGAN, which is a data-driven DL-based approach, aims to generate noise samples with complicated noise statistics that can not be easily captured using existing noise models. Simulation results confirmed that the proposed NGGAN can generate noise samples with statistics similar to those of the original dataset while maintaining a certain level of diversity.

From a measurement perspective, the proposed NGGAN can be used to model complex noise traces for NB-PLC systems and may prove to be a more efficient learnable data augmentation method than traditional model-based methods. This offers significant advantages in training DL-based NB-PLC transceivers~\cite{Tonello2019}, which consider the impact of complicated noise statistics of the noises for NB-PLC systems~\cite{Lemley2020}. This concept can be extended to the development of consumer electronic devices that suffer from complicated noise~\cite{Sharma2019}.

The main contributions of this work are summarized as follows:
\begin{itemize}
\item To closely match the statistics of complicated noise over NB-PLC systems, we measured the NB-PLC noises via the analog coupling and bandpass filtering circuits of a commercial NB-PLC modem to build a realistic dataset~\cite{ChienDataset2023} from different scenarios involving fans, lamps, and power suppliers~\cite{Antoniali2014}.
\item To extract the features of the NB-PLC noise waveform precisely, the length of the input data was determined based on cyclo-stationary properties.
\item To enhance the similarity between the generated noise and training, the Wasserstein distance was used to replace the Kullback–Leibler (KL) divergence as the loss function of the original GAN to enhance the similarity between the generated noise and three datasets: (i) PSCGM, (ii) FRESH, and (iii) practical measurements from an NB-PLC network (the three datasets are available on the IEEE DataPort~\cite{ChienDataset2023}).
\item To measure the similarity performances of GAN-based models, several performance metrics are adopted, including (1) maximum value, (2) mean value, (3) energy value, (4) standard deviation, (5) skewness, (6) kurtosis, (7) the number of peaks larger than the certain threshold value, (8) skewness of autocorrelation, (9) kurtosis of autocorrelation, (10) cyclic spectral density (CSD), (11) cyclic spectral coherence (CSC), (12) principal component analysis (PCA), and (13) Fréchet inception distance (FID).
\item Simulation results demonstrate that the proposed NGGAN is a more efficient data augmentation approach than other GAN-based models for improving its robustness against noise~\cite{Lemley2020} for the NB-PLC transceiver design.
\item The Python source codes are provided on GitHub\footnote{\url{https://github.com/yrchien/NGGAN}}.
\end{itemize}

The remainder of this paper is organized as follows:
Section~\ref{sec:related} introduces related work on the modeling of NB-PLC noise.
Section~\ref{sec:proposed} describes the proposed NGGAN model.
Section~\ref{sec:eva} outlines the performance metrics, followed by a presentation of the datasets and noise measurements in the NB-PLC networks.
Section~\ref{sec:sim} presents qualitative and quantitative analyses of the noise generated by various GAN models developed for NB-PLC systems.
The conclusions are summarized in Section~\ref{sec:conclusions}.

\section{Related Works}
\label{sec:related}
In this section, we review related studies on NB-PLC noise models with cyclo-stationary properties.
First, the PSCGM was adopted as the standard noise model under IEEE 1901.2~\cite[Annex D.3]{IEEE1901}.
Second, the FRESH filter, which is a parametric approach, is presented~\cite{Moaveninejad2020}.
Finally, GAN-based methods for generating NB-PLC noise were examined.

\subsection{PSCGM Modeling}
The IEEE standard 1901.2~\cite{IEEE1901} proposed the PSCGM model, which includes three main components: random background noise, periodic impulsive noise, and random impulse noise based on field measurements over low-voltage (LV) sites.
The PSCGM divides each period of the cyclo-stationary noise into two or three regions to which specific temporal shaping and spectral shaping filters are assigned~\cite{Hayes1996}.

However, the PSCGM is measured using two-dimensional amplitude spectrograms of cyclo-stationary noise to visually distinguish different regions~\cite{Hayes1996}, such as the number and size of pulses. Therefore, the correlation of Gaussian noise between different regions is not considered.

\subsection{FRESH Filter-based Modeling}
\label{subsec:FRESH}
The noise traces generated by a set of FRESH filters~\cite{Elgenedy2016} can be categorized into three classes based on the standard deviation of one slot.
A set of parameterized spectral and temporal shaping filters~\cite{Moaveninejad2020} was provided for the generated cyclo-stationary noise using white Gaussian noise as the excitation input.
The input was first transformed into the frequency domain using a fast Fourier transform (FFT) to undergo shaping using two asymmetric double-sided exponential decay functions (i.e., spectral shaping filters).
The frequency-shaped signal was subjected to an inverse FFT to obtain a sequence in the time domain, which was subjected to temporal shaping using a symmetric double-sided exponential decay function.

Typically, the assessment of the FRESH filter-based model employs a normalized mean square error (MSE). An effective strategy for reducing the normalized MSE is to augment both the quantity and size of the FRESH filters. Nonetheless, implementing the FRESH model proved challenging because of the substantial number of parameters associated with each filter. Consequently, the increased complexity of the FRESH model results in an extended duration of noise generation, as highlighted in~\cite{Moaveninejad2020}.

\subsection{GAN-based Modeling}
\label{sec:GAN_Modeling}

The GAN~\cite{Goodfellow2014} is widely used to extract features with structural properties within a dataset and to synthesize samples with statistical properties similar to those of training samples~\cite{Radford2015, Gulrajani2017, Donahue2018, Yoon2019, Letizia2020, Chien2021, Stanczuk2021}.
An effective GAN model for time-series data should maintain temporal dynamics to ensure that the generated sequences uphold the original relationships between variables over time~\cite{Yoon2019}.
However, it was difficult to converge for the training GAN until the advent of unsupervised learning in the form of deep convolutional GANs (DCGANs)~\cite{Radford2015}, which are characterized by the use of a convolutional layer in the generator and a convolutional-transpose layer in the discriminator.
The DCGAN converges more easily than the original GAN, whereas the batch normalization layer enhances stability.
In~\cite{Letizia2020}, the short-time Fourier transform (STFT) of noise traces is transformed by DCGAN into a two-dimensional amplitude spectrogram. 
Then, the Griffin-Lim algorithm~\cite{Griffin1984} is applied to transform the two-dimensional spectrogram into a one-dimensional time-domain sequence. 
However, the Griffin-Lim algorithm leads to phase deviations in the time-domain sequence, such that the statistical characteristics differ considerably from the practical NB-PLC noise.
This leads to poor performance of the DCGAN in generating cyclo-stationary noise~\cite{Gulrajani2017}.
The spectrogram-based GAN (SpecGAN) further improves the DCGAN performance using the Wasserstein distance~\cite{Gulrajani2017, Stanczuk2021} as the loss function.

In our previous study~\cite{Chien2021}, we proposed a novel end-to-end GAN model called the phase-learned SpecGAN (PL-SpecGAN) to account for the phase spectrum without using phase estimation.
The PL-SpecGAN further incorporates phase data related to noise signals within a SpecGAN to simultaneously combine the phase and amplitude information simultaneously.
It also resolves phase deviation problems using the Griffin-Lim algorithm for inverse STFT operation.
When dealing with periodic data, the learning performance of a GAN can be improved by increasing the length of the training data.
Inspired by~\cite{Donahue2018}, we reduced the dimension of the convolutional layer from two to one and extended the length of the feature filter for the proposed NGGAN.
On the other hand, we adopted the frequency-domain version of SpecGAN (FD-SpecGAN)~\cite{Radford2015} with modifications to the architecture of the generative and discriminative models to fit our previous framework~\cite{Chien2021}.
We compare the performance of the DCGAN, PL-SpecGAN, FD-SpecGAN, and NGGAN using three types of datasets.

\section{Proposed NGGAN and Training Processing}
\label{sec:proposed}
\begin{figure}[tbh]
\centering
\includegraphics[width=0.4\textwidth]{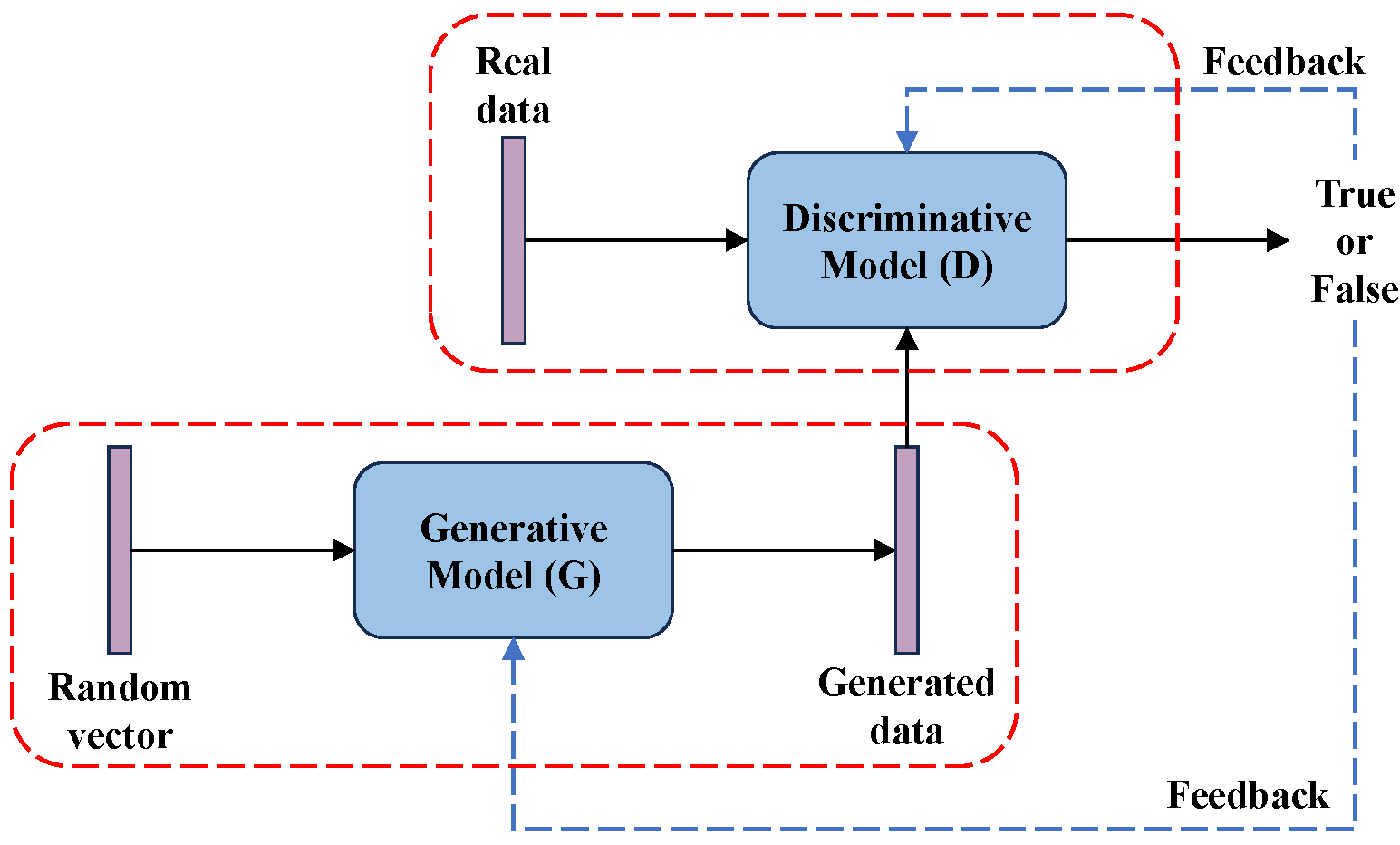}
\caption{The GAN-based architecture.}
\label{fig:GAN_generator}
\end{figure}

In this section, the proposed NGGAN algorithm and its training processes are introduced. 
Fig.~\ref{fig:GAN_generator} depicts the architecture of the GAN-based model. A GAN comprises a generative (G) model, which is responsible for capturing the data distribution, and a discriminative (D) model, which determines whether the samples are training data or data generated from the G model.

\begin{figure}[tb]
    \subfloat[\label{subfig:G_network}]{%
      \includegraphics[width=0.45\textwidth]{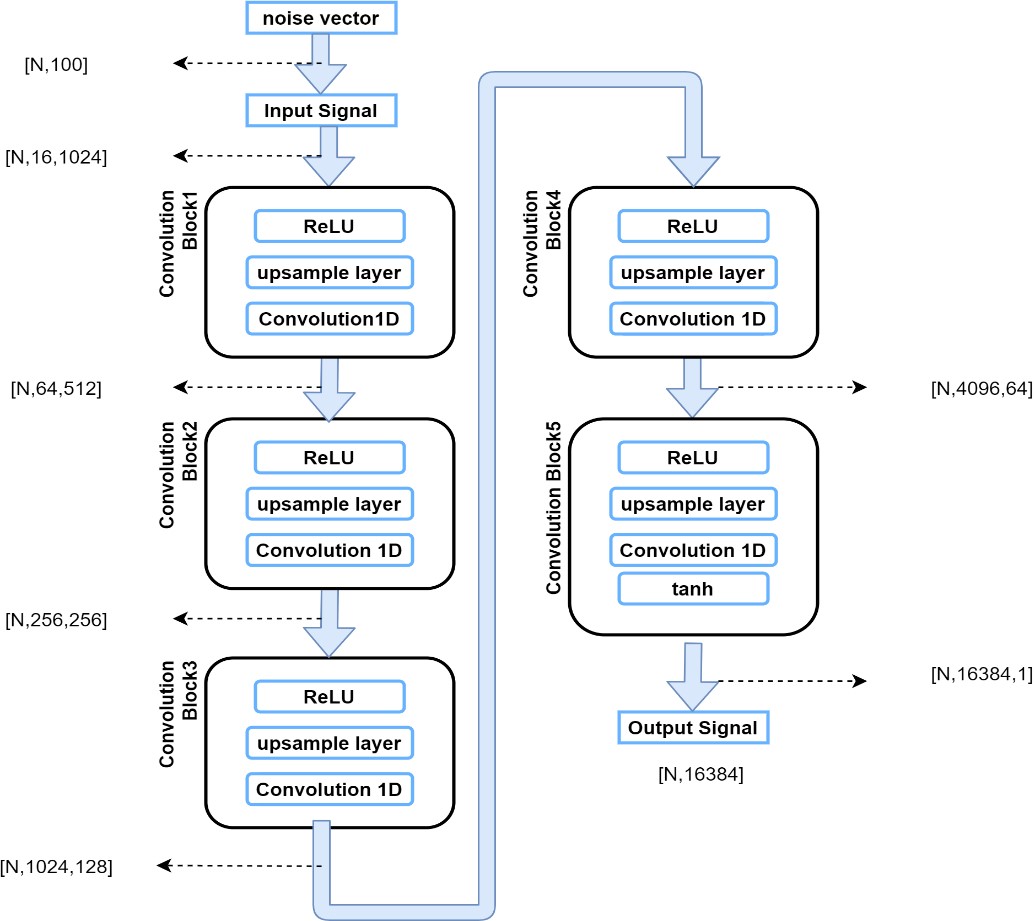}
    }
    \hfill
    \subfloat[\label{subfig:D_network}]{%
      \includegraphics[width=0.45\textwidth]{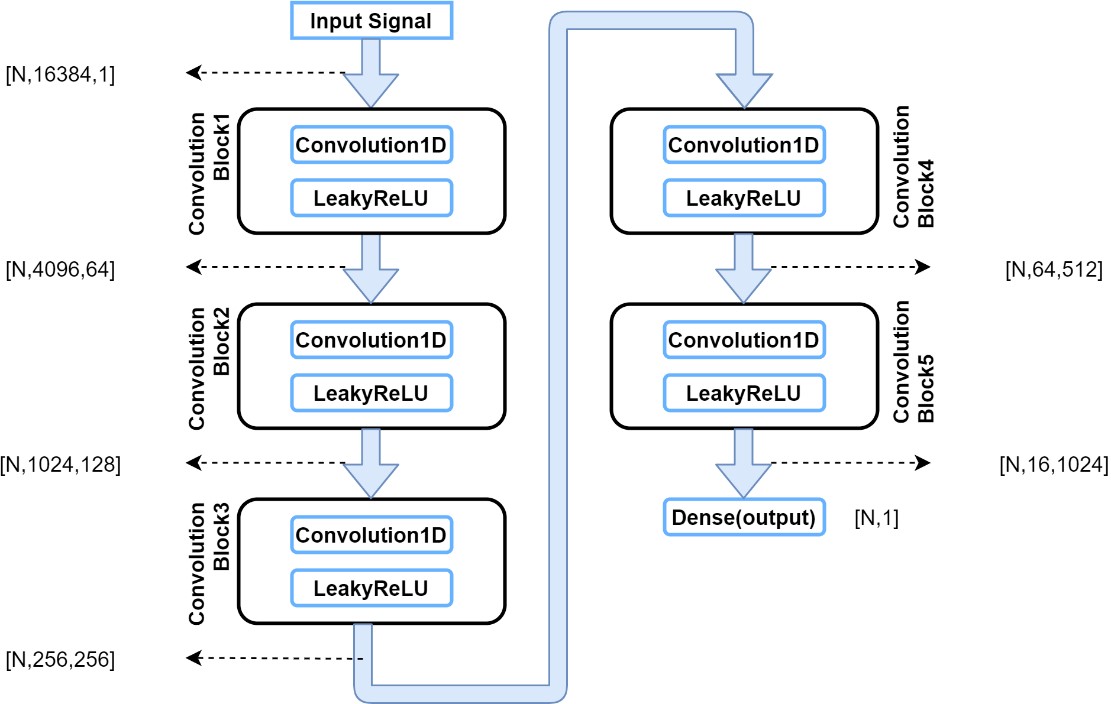}
    }
   \caption{The proposed NGGAN architecture: (a) generator and (b) discriminator ($N$ is the total number of impulse noise traces).}
\label{fig:GAN T.G. Structure}
  \end{figure}

Fig.~\ref{fig:GAN T.G. Structure} illustrates the proposed NGGAN architecture.
The parameter settings for each layer are listed in Table~\ref{tab:GAN T.G. parameters}, where $N$ denotes the batch size.
In the one-dimensional convolutional (Conv1D) layer, the three-tuple parameter in the field ``filter size" respectively indicates the length of the filter, the stride, the length of input data, and the depth of the filter.
The three-tuple parameter in the ``output size" denotes the batch size. 
The length of the feature filter was set to $25$, based on an expanded two-dimensional $5 \times 5$ feature filter to observe the out-of-range signals in our previous work~\cite{Chien2021}, where the length of the feature filter was set to $5$.
This enhances the precision of learning the correlation and periodicity between noise sequences.

\subsection{Generative Model}
\label{subsec:net_design}
As shown in Fig.~\ref{fig:GAN T.G. Structure} (a), the proposed generator comprises five concatenated convolutional blocks, each of which comprises an upsampling layer with a rectified linear unit (ReLU) activation function.
The details of the feature filter parameters are listed in Table~\ref{tab:GAN T.G. parameters} (a).
The input noise vector (length = $100$) is sampled from a uniformly distributed random variable with a range of $[-1, 1]$.
In a previous study, we adopted a 1D transposed convolutional layer for learning~\cite{Chien2021}.

In this study, we increased the number of layers and the length of the generated data $4$ times, resulting in a generation vector with a length of $16,384$.
The block number of the generator and discriminator depends on the application of the GAN model.
Periodic stationary impulse noise occurs with a frequency of approximately half the alternating current (AC) cycle~\cite{Cooper2002}, for example, 8.33 ms is a cycle of impulse noise at an AC frequency of 60 Hz. 
If the sampling period is $2.5\mu$ seconds (equivalent to $1/400$ kHz), then $16,384$ samples approximately five cycles of impulse noise.
This allows the proposed NGGAN to capture features or correlations that extend for up to five cycles.
The depth of the feature filter was decreased from $512$ to $1$.

In our previous study~\cite{Chien2021}, the simulation results demonstrated that a transposed convolutional layer degraded the upsampling learning performance because most of the interpolation values were $0$.
In this paper, we propose a novel upsampling method that combines the nearest-neighbor algorithm and linear interpolation to allow the insertion of the same value in adjacent areas, thereby enlarging the signal length (by $4 \times$) to preserve more information.
The stride was set to $1$ for the ConV1D layer.
The ReLU was adopted as the activation function before each transposed convolutional layer.
The hyperbolic tangent function was adopted as the activation function for the output layer to normalize values in the time-domain sequence $[-1,1]$.
A fully connected layer is used to map the noise vector onto the input vector with a length of $16,384$.
\subsection{Discriminative Model}
As shown in Fig.~\ref{fig:GAN T.G. Structure} (b), the discriminator includes five convolutional layers and a leaky ReLU with a parameter set to $0.2$.
The details of the feature filter parameters are listed in Table~\ref{tab:GAN T.G. parameters} (b).
By setting the stride to 4, the feature filter reduces the signal length by $4\times 4$ layer-by-layer, such that the length of the noise vector in the first convolutional layer is reduced from $16,384$ to $4,096$.
Thus, an eigenvector length of $16$ was obtained using five convolutional layers.
The filter depth was doubled in each consecutive layer to allow the filter to learn comprehensive features from the precise features.
The output vector of the last layer had a size of $[16,1024]$, and the data were expanded into a vector with a length of $16,384$.
The features were extracted by the filter, and weights were assigned via the fully connected layer to obtain the optimal value learned by the discriminator. 
After each convolution, the leaky ReLU activation function had a negative slope of $0.2$ after each convolution.

\begin{table}[t]
\centering
\caption{Detailed parameters of each layer in (a) generator and (b) discriminator networks.}
\scalebox{0.85}{
 \subfloat[][]{
      \begin{tabular}{|c|c|c|c|}
      \hline
        \textbf{Layer name} & \textbf{Filter size} & \textbf{Activation function} & \textbf{Output size} \\
        \hline
        \textbf{Noise vector} & (100, 16384) & - & (N, 100)\\\hline
        \textbf{Input} & - & - & (N, 16, 1024)\\\hline
        \textbf{Conv1D} & (25, 1, 1024, 512) & ReLU & (N, 64, 512)\\\hline
        \textbf{Conv1D} & (25, 1, 512, 256) & ReLU & (N, 256, 256)\\\hline
        \textbf{Conv1D} & (25, 1, 256, 128) & ReLU & (N, 1024, 128)\\\hline
        \textbf{Conv1D} & (25, 1, 128, 64) & ReLU & (N, 4096, 64)\\\hline
        \textbf{Conv1D} & (25, 1, 64, 1) & ReLU & (N, 16384, 1)\\\hline
        \textbf{Output} & - & tanh & (N, 16384)\\
        \hline
       \end{tabular}}
       }
\scalebox{0.85}{
  \subfloat[][]
  {
         \begin{tabular}{|c|c|c|c|}
         \hline
        \textbf{Layer name} & \textbf{Filter size} & \textbf{Activation function} & \textbf{Output size} \\
        \hline
        \textbf{Input} & - & - & (N, 16384, 1)\\\hline
        \textbf{Conv1D} & (25, 4,1, 64) & Leaky ReLU(0.2) & (N, 4096, 64)\\\hline
        \textbf{Conv1D} & (25, 4, 64, 128) & Leaky ReLU(0.2) & (N, 1024, 128)\\\hline
        \textbf{Conv1D} & (25, 4, 128, 256) & Leaky ReLU(0.2) & (N, 256, 256)\\\hline
        \textbf{Conv1D} & (25, 4, 256, 512) & Leaky ReLU(0.2) & (N, 64, 512)\\\hline
        \textbf{Conv1D} & (25, 4, 512, 1024) & Leaky ReLU(0.2) & (N, 16, 1024)\\\hline
        \textbf{Dense(Output)} & (16384, 1) & - & (N, 1)\\
        \hline
       \end{tabular}  
  }
  }
     \label{tab:GAN T.G. parameters}
\end{table}

\subsection{Data Pre-processing}
Effective training data pre-processing and parameter initialization are essential or enhancing the efficiency and accuracy of the trained model. 
By employing data pre-processing techniques, we can obtain the following two advantages. (i) The first benefit is convergence speed enhancement. 
This causes the shape of the loss function to become narrow and elongated, resulting in longer iteration times. Data pre-processing techniques can significantly improve the convergence speed. (ii) The second benefit is improved model accuracy.
There are three data pre-processing techniques, including: (1) min-max normalization, (2) Z-score standardization, and (3) PCA.
Employing these three data pre-processing techniques, the subsequent stage is the selection of an appropriate loss function. The role of the loss function is to provide continuous feedback during the training process in the selected model. This feedback empowers the model to adjust its parameters and improve its performance for a given task.

\section{Performance Metrics and Training Datasets}
\label{sec:eva}

\subsection{Performance Metrics}
\label{sec:performance_metrics}
Let $s_i(n)$ be the $n$-th samples in the $i$-th generated trace, each of which includes $N$ samples $(n=1, 2, \ldots, N)$.
We evaluated the quality of the generated noise samples using the following statistical items:\\
    1) Maximum value:
    \begin{align}
        m_i = \underset{n}{\textrm{max}}\;s_i[n].
    \end{align}
    2) Mean value: 
    \begin{align}
        \mu_i = \frac{1}{N}\sum_{n=1}^{N}s_i[n].
    \end{align}
    3) Energy value:
    \begin{align}
        P_s = \frac{1}{N}\sum_{n=1}^Ns_i^2[n].
    \end{align}
    4) Standard deviation of the time-domain sequence:
    \begin{equation}
        \sigma_s = \sqrt{\frac{1}{N-1}\sum_{n=1}^{N}(S_i[n]-\mu_s)^2}.
    \end{equation}     
    5) Skewness: The asymmetry of a probability distribution is calculated by 
    \begin{equation}
        S_s = \frac{ \frac{1}{N}\sum_{n=1}^{N} (S_i[n]-\mu_{s})^3 }{(\frac{1}{N}\sum_{n=1}^N (S_i[n]-\mu_{s})^2)^\frac{3}{2}},
        \label{eq:skewness}
    \end{equation}     
    where positive/negative skewness indicates that the probability density is skewed toward the right or left.\\
    6) Kurtosis: The peak of the probability distribution is calculated as 
    \begin{equation}
        K_s = \frac{ \frac{1}{N}\sum_{n=1}^N (S_i[n]-\mu_{s})^4 }{(\frac{1}{N}\sum_{n=1}^N (S_i[n]-\mu_{s})^2)^2},
        \label{eq:kurtosis}
    \end{equation}     
    where a high peak indicates that the variance is increased by extreme outlier values.\\
    7) Number of peaks over 0.05V: The number of peaks in a sample exceeding 0.05V is calculated by
    \begin{equation}
        NP_s = \#\{ \left|S_i[n]\right| > 0.05\},
    \end{equation}
    where $\#\{\cdot\}$ denotes the counting function.\\    
    8) Skewness of autocorrelation: The autocorrelation sequence is defined as $r_k=\frac{c_k}{c_0}$, where $c_k$ is defined as $c_k=\frac{1}{N}\sum_{n=1}^{N-k}(s_i[n]-\mu_s)(s_i[n+k]-\mu_s)$. By substituting the autocorrelation sequence $r_k$ and its mean $\mu_r$ into~\eqref{eq:auto-correlation_s}, the sample autocorrelation skewness value $SA_s$ is obtained by
    \begin{equation}
        SA_s = \frac{ \frac{1}{N}\sum_{k=1}^{N} (r_k-\mu_{r})^3 }{(\frac{1}{N}\sum_{k=1}^N (r_k-\mu_{r})^2)^\frac{3}{2}}.
        \label{eq:auto-correlation_s}
    \end{equation}   
    9) Kurtosis of autocorrelation: By substituting the autocorrelation sequence $r_k$ and its mean $\mu_r$ into~\eqref{eq:auto-correlation_k}, the autocorrelation peak value $KA_s$ is obtained by
    \begin{equation}
        KA_s = \frac{ \frac{1}{N}\sum_{k=1}^N (r_k-\mu_{r})^4 }{(\frac{1}{N}\sum_{k=1}^N (r_k-\mu_{r})^2)^2}.
        \label{eq:auto-correlation_k}
    \end{equation}
   10) CSD and CSC: The autocorrelation of cyclo-stationary noise exhibits periodic characteristics and can be estimated from cyclic autocorrelation and cyclic power spectral density to identify the original cyclo-stationary noise~\cite{Elgenedy2016}. 
    Therefore, we employed CSD and CSC plots to individually examine how the intensity components of the noise changed at different frequencies and their correlations.
The cyclic autocorrelation function (CAF), which is expressed as follows:
\begin{align}
    R^{\alpha_k}[\tau]=\frac{1}{P}\sum_{n=0}^{P-1}r_{k}[n;\tau]e^{-j2 \pi \alpha_k n},
\end{align}
where $\alpha_k=k/P$, $k=0, 1,\ldots, P-1$, denotes the $k$-th cyclic frequency of $r_{k}[n;\tau]$.
The CSD function is calculated as follows:
\begin{align}
   S[\alpha_k;f]=\sum_{\tau=-\infty}^{\infty} R^{\alpha_k}[\tau]e^{-j 2 \pi f \tau},
   \label{eq:CSD}
\end{align}
where $f$ denotes the spectral frequency.
The normalized CSD, referred to as the CSC function, is defined as follows:
\begin{align}
   \overline{S}[\alpha_k;f]=\frac{S(\alpha_k;f)}{\sqrt{S(0;f)S(0;f+\alpha_k)}}.
   \label{eq:CSC}
\end{align}
11) PCA and FID: When $15,000$ data samples were generated, each sample comprised eight features, where statistical features exceeding $0.05$ V peaks were excluded as the primary component features for each data. The derivation of PCA is as follows. We consider a matrix $\mathbf{A}$ with dimensions of $15000\times8$, where $\mathbf{A}_{ij}$ denotes the element in the $i$-th row and $j$-th column. The formula for covariance matrix $\mathbf{cov}_{kl}$ is given by
\begin{equation}
    \mathbf{cov}_{kl} = \sum_{i=1}^{15000}(\mathbf{A}_{ik}-\mu_{A_k})(\mathbf{A}_{il}-\mu_{A_l}),
\end{equation}
where $\mu_{A_k}=\frac{1}{15000}\sum_{i=1}^{15000}\mathbf{A}_{ik}$, and $k,l\in \{1, 8\}$. Then, we can obtain the eigenvector matrix $\mathbf{X}\in \mathbb{R}^{8\times 8}$, and $\mathbf{X}_{ij}$ denotes the $i$-th element in the $j$-th eigenvector. Projecting $\mathbf{A}$ onto the eigenvector matrix $\mathbf{X}$ yields the projection matrix $\mathbf{Y}_{qp}= \sum_{i=1}^{8}\mathbf{X}_{ip}\mathbf{A}_{qi}$, where $q=1,2,\ldots,15000$, and $p=1,2,\ldots,8$.
The PCA scatter reveals the proximity of the synthetic data distribution to real data in a 2-dimensional space and indicates whether the generated data adequately spans the area of the original data.
Discrepancies between the training and generated data are treated as diversity.
This facilitates fidelity and diversity of the generated data~\cite{Yoon2019}.

The FID value for the noise generated by each GAN-based model is calculated as
\begin{equation}
    \text{FID}(x,g) = ||\mu_x-\mu_g||_2^2 + {\rm Tr}(\mathbf{\Sigma}_x+\mathbf{\Sigma}_g-2(\mathbf{\Sigma}_x\mathbf{\Sigma}_g)^{1/2}),
    \label{eq:FID}
\end{equation}
where $\mu_x$ and $\mu_g$ represent the mean values of the principal component eigenvectors in the training and generated sets, respectively, and $\mathbf{\Sigma}_x$ and $\mathbf{\Sigma}_g$ refer to the covariance matrices derived from the principal component eigenvectors of the training and generated sets, respectively. A lower FID value indicates better quality and diversity of the generated data (closer to the training data distribution).
Among all the performance metrics, the PCA scatter and FID values stand out as the most critical because they effectively represent the fidelity and diversity of the generated samples.

\subsection{Training Datasets}
\label{sec:dataset}

\begin{figure*}[tb]
\centering
    \subfloat[\label{subfig:dataset1}]{%
      \includegraphics[width=0.33\textwidth]{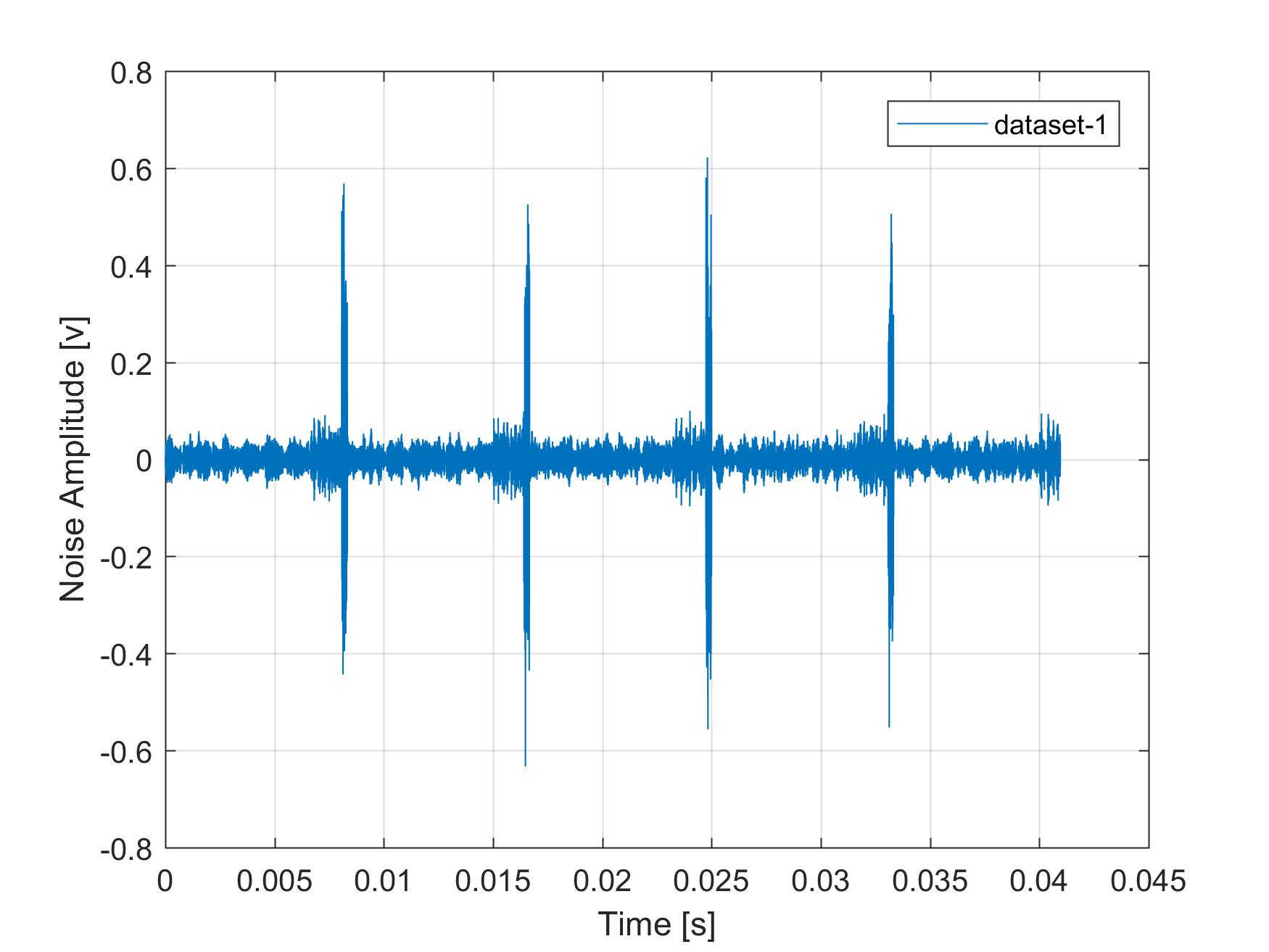}
    }
    \subfloat[\label{subfig:dataset2}]{%
      \includegraphics[width=0.33\textwidth]{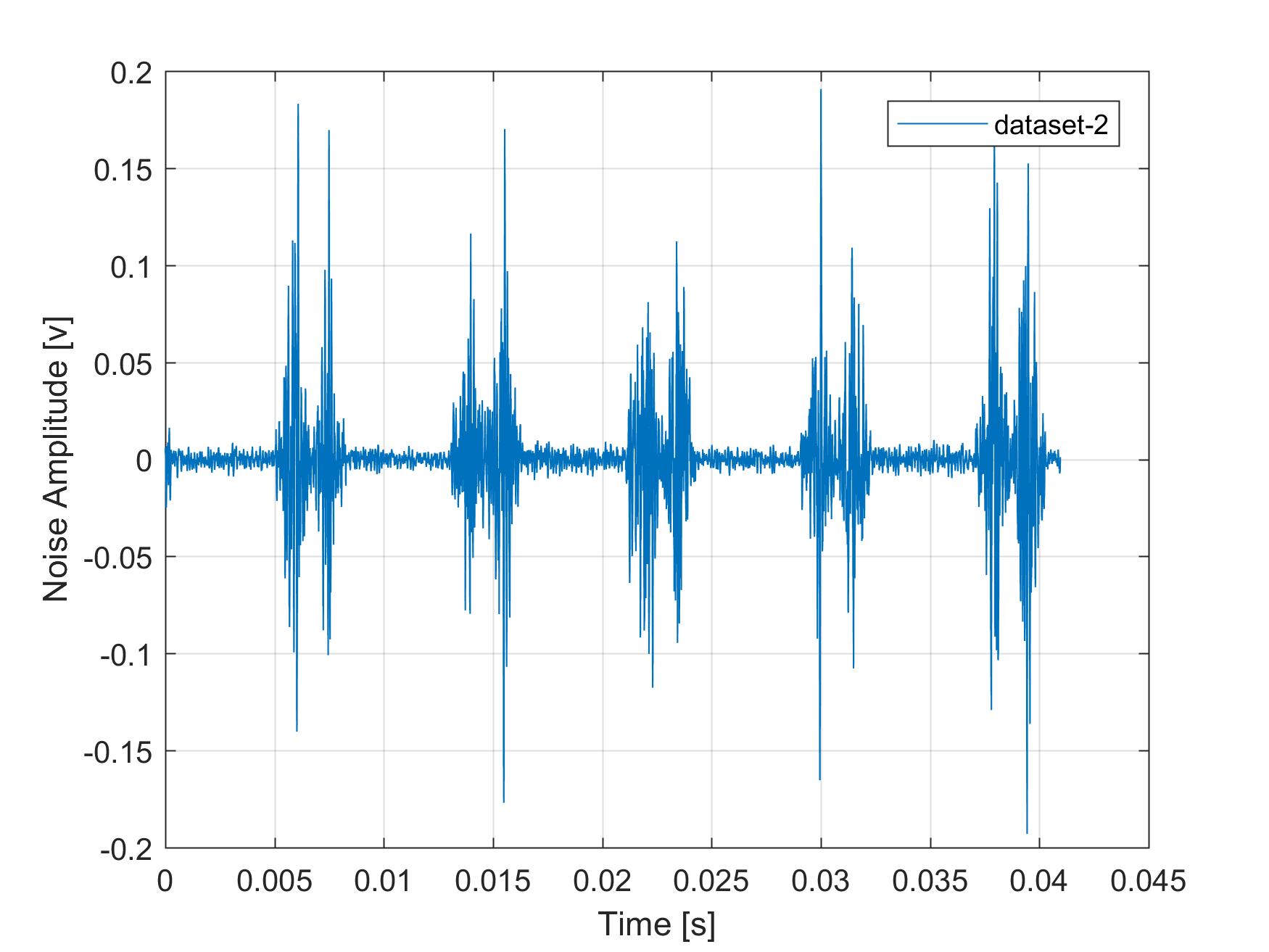}
    }
    \subfloat[\label{subfig:dataset3}]{%
      \includegraphics[width=0.33\textwidth]{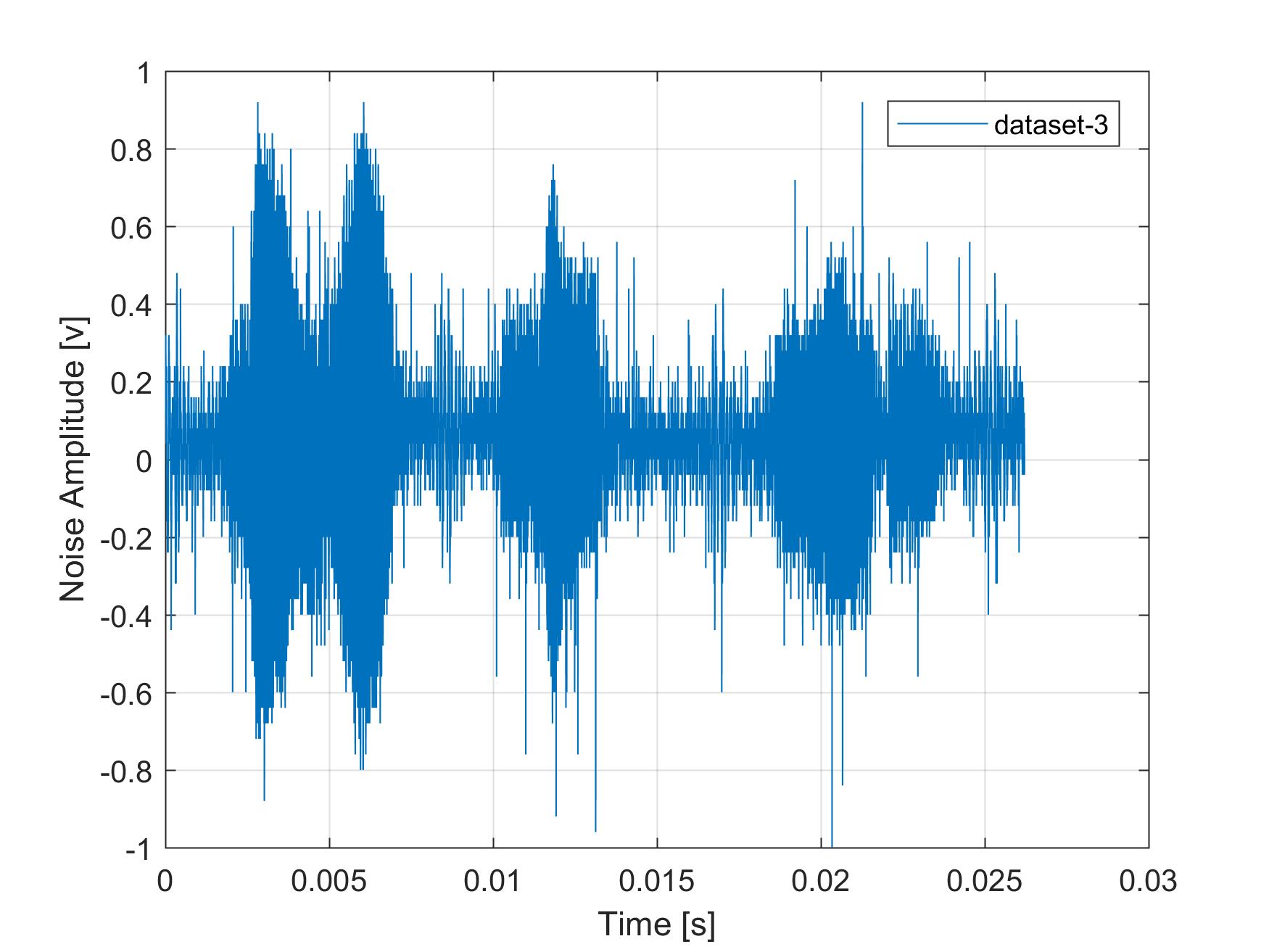}
    }
   \caption{Illustration of sample traces for each dataset: (a) Dataset-1, (b) Dataset-2, and (c) Dataset-3.}
\label{fig:dataset}
  \end{figure*}
  
Using these three datasets, we compared the accuracy and diversity of the GAN-based models.
Each dataset included $15,000$ records comprising $16,384$ samples ($40.96$ms) and is publicly accessible~\cite{ChienDataset2023}.
The first dataset is a PSCGM-generated noise trace obtained using the parameters outlined in LV14~\cite[Annex 14]{IEEE1901}, as illustrated in Fig.~\ref{fig:dataset} (a).
The second dataset is a FRESH-generated noise~\cite{Moaveninejad2020}, as illustrated in Fig.~\ref{fig:dataset} (b).
The third dataset is the real impulse noise collected using an analog coupling circuit at the front end of a commercial power-line modem development kit with a sampling rate of 625 kHz, as illustrated in Fig.~\ref{fig:dataset} (c).

\begin{figure}[tb]
\centering
\includegraphics[width=0.5\textwidth]{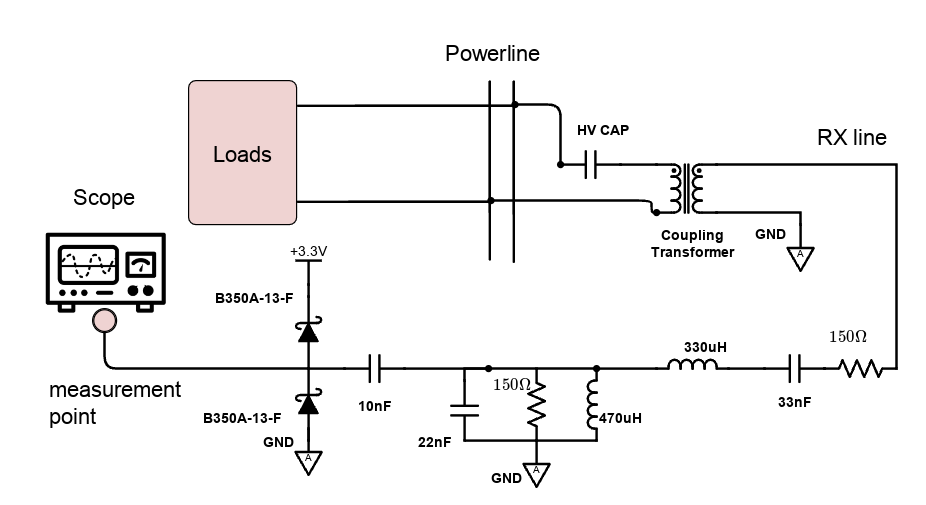}
\caption{The coupling circuit and analog bandpass filter in our measurement~\cite{TI2014}.}
\label{fig:TIDM-TMDSPLCKIT-V3}
\end{figure} 

\begin{figure}[tb]
\centering
\includegraphics[width=0.4\textwidth]{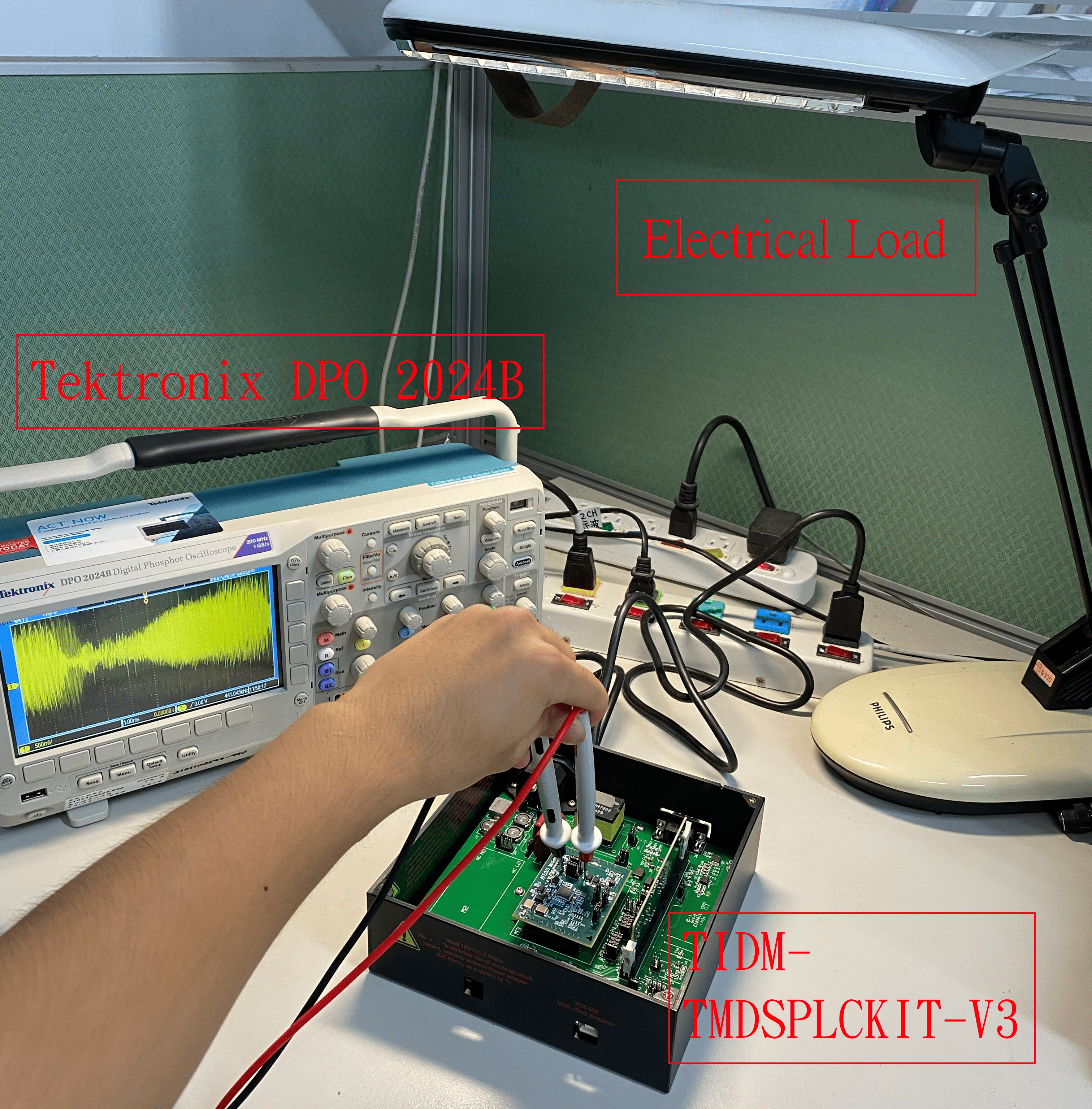}
\caption{Collecting the NB-PLC noise samples for Dataset-3.}
\label{fig:measuring}
\end{figure}

Fig.~\ref{fig:TIDM-TMDSPLCKIT-V3} shows the circuits that the Texas Instruments (TI) PLC Developer’s Kit TIDM-TMDSPLCKIT-V3 used to measure the noise for Dataset-3.
To prevent the measurement circuit from being damaged, the coupling circuit is used to block the 110V/220V mains component while the powerline channel is transmitting and receiving~\cite{Artale2013}.
In addition, the coupling circuit is used to ensure impedance matching of the measurement circuit and to avoid the impact of voltage spikes or fast electrical transient (burst) pulses on the measurement circuit.
When the PLC signal from the coupling circuit passes through a fourth-order passive bandpass filter (24–105 kHz) via RX Line, the NB-PLC noise is measured using a Tektronix DPO 2024 B digital oscilloscope with a sampling frequency of 625 kHz.

As shown in Fig.~\ref{fig:measuring}, the electrical load~\cite{Antoniali2014} is connected to a powerline to generate cyclic pulses. 
A total of $2.4576\times10^8$ sample points were collected as one piece of noise data.
The total number of noise data records is 15,000, and the length of each noise data sample is 16,384.

\begin{figure*}[tb]
\centering  
\includegraphics[width=0.95\textwidth]{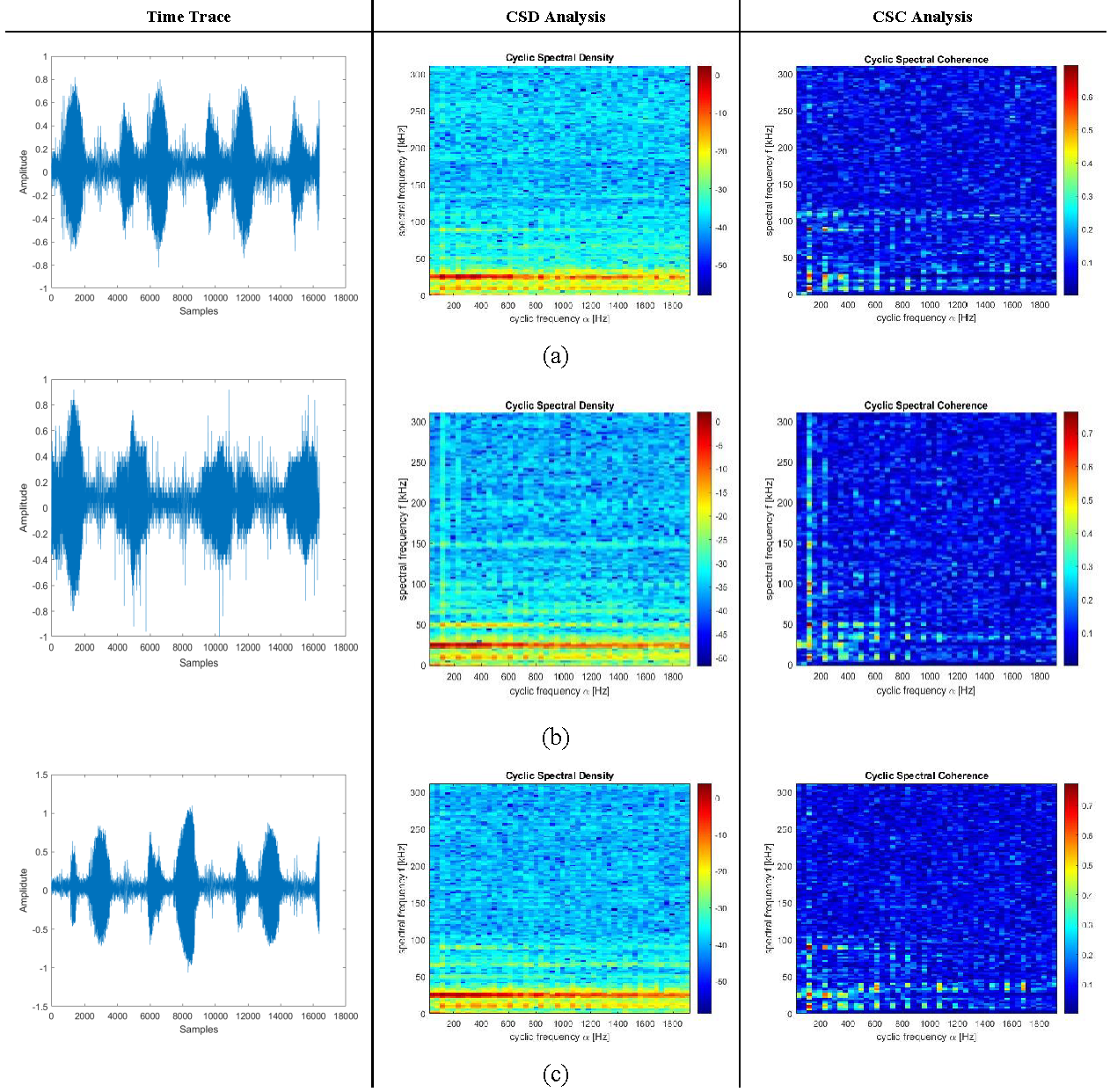}
  \caption{Examples of various loading used when measuring noise in the NB-PLC systems: (a) Fans, (b) lamps, and (c) power suppliers.}
  \label{fig:dataset3-example}
\end{figure*}

When collecting the measurement data for Dataset-3, we considered a wide range of scenarios to ensure that Dataset-3 includes a variety of noise sample types with diverse noise characteristics. This encompassed variations in loading within the powerline network and different powerline network topologies. For example, we illustrate three scenarios involving fans, lamps, and power supplies~\cite{Antoniali2014} (as shown in Fig.~\ref{fig:dataset3-example}). The waveforms exhibited distinct sharpness. For a more comprehensive analysis, CSD and CSC were adopted.
Note that the Dataset-3 trajectories encompassed diverse scenarios, including loading across various powerline network topologies.
Therefore, Dataset-3 was more complex than the samples generated in Dataset-1 and Dataset-2.
The three datasets are available in the IEEE DataPort~\cite{ChienDataset2023}.

\section{Simulation Results}
\label{sec:sim}

For the three datasets described in Sec.~\ref{sec:dataset}, we evaluate the 13 performance metrics mentioned in Sec.~\ref{sec:performance_metrics} for the four GAN-based models: DCGAN~\cite{Letizia2020}, FD-SpecGAN~\cite{Radford2015}, PL-SpecGAN~\cite{Chien2021}, and the proposed NGGAN in Sec.~\ref{sec:GAN_Modeling}.
The hyperparameter settings of the four GAN-based models are as follows: learning rate $10^{-4}$, epochs 200, and batch size 64 during the training process.
The generated samples were examined qualitatively and quantitatively to assess the performance of the GAN-based models in emulating the cyclo-stationary pulse noise.
To enhance the convergence rate of the proposed NGGAN model, we utilized four techniques: a batch normalization layer, dropout, L2 regularization, and early stopping.
The Python source code for this work can be found on GitHub\footnote{\url{https://github.com/yrchien/NGGAN}}, including comprehensive details about parameter settings.

\subsection{Dataset-1 (PSCGM-Generated)}

\begin{figure}[tb]
\centering
  \includegraphics[width=0.45\textwidth]{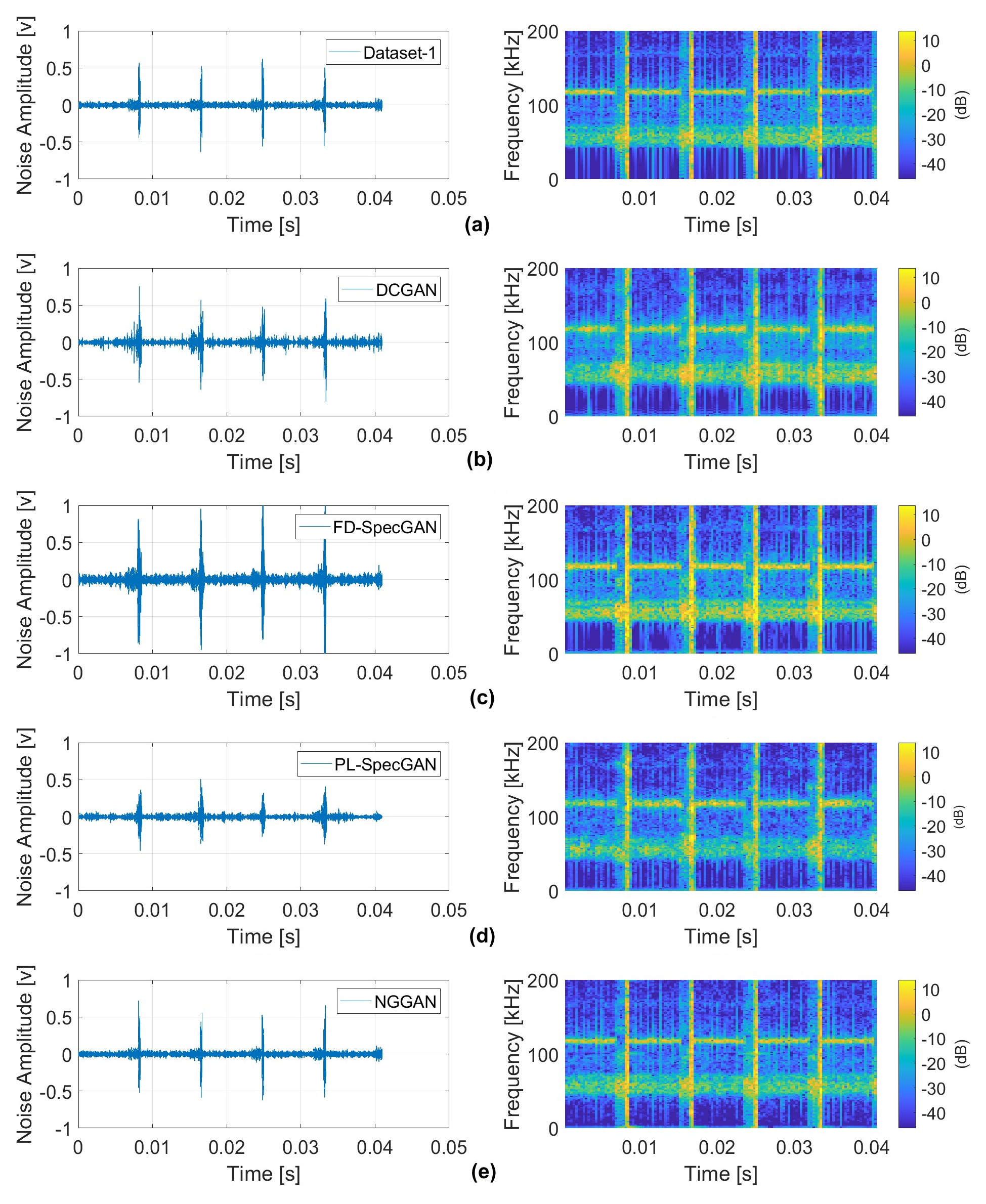}
  \caption{Noise time series and spectrograms: (a) Dataset-1; (b) DCGAN~\cite{Letizia2020}; (c) FD-SpecGAN~\cite{Radford2015}; (d) PL-SpecGAN~\cite{Chien2021}; (e) NGGAN.}
  \label{fig:trace_spectrogram_1}
\end{figure}

Fig.~\ref{fig:trace_spectrogram_1} presents the noise time series and corresponding spectrograms for the four GAN-based models learning Dataset-1. 
The spectrograms reveal that the frequency components of the generated noise varied periodically with time, as did the time series. 
The proposed NGGAN outperformed the other GAN-based models in terms of the mean, standard deviation, and median feature statistics. 
The NGGAN outperformed the DCGAN in terms of maximum values (+9$\%$), energy values (+34$\%$), and standard deviation (+15$\%$), where feature values are represented as 100$\%$ and higher values indicate greater similarity. 
For example, if the feature value of the NGGAN-generated noise is 98$\%$ (a difference of 2$\%$) and that of the DCGAN-generated noise is 88$\%$ (a difference of 12$\%$), then the improvement afforded by the NGGAN would be 10$\%$ (12$\%$-2$\%$).
In addition, PL-SpecGAN slightly outperformed FD-SpecGAN because it did not use the Griffin-Lim process for the loss estimation.

\begin{figure}[tb]
  \centering
  \includegraphics[width=.45\textwidth]{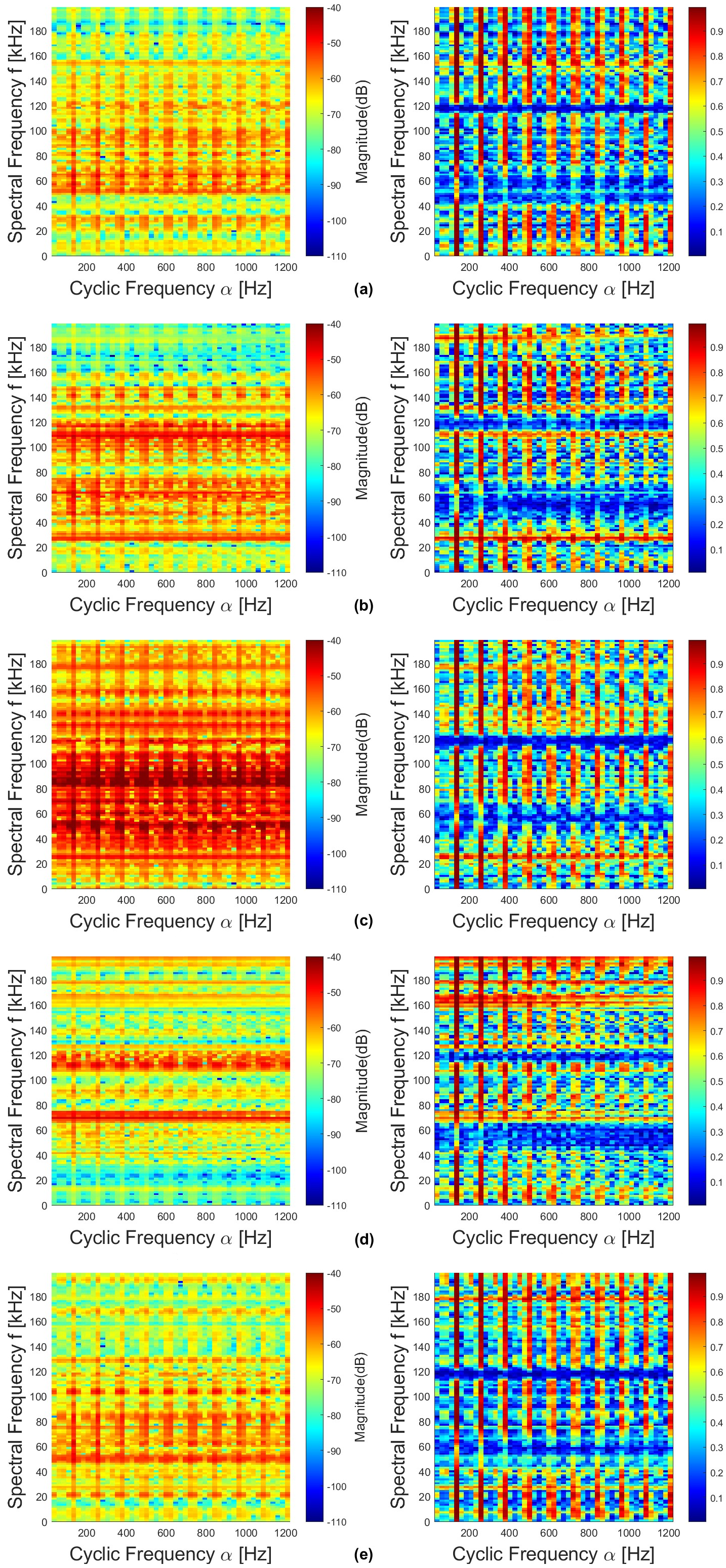}
  \caption{CSD and CSC graphs: (a) Dataset-1; (b) DCGAN~\cite{Letizia2020}; (c) FD-SpecGAN~\cite{Radford2015}; (d) PL-SpecGAN~\cite{Chien2021}; (e) NGGAN.}
  \label{fig:total1_cyclic}
\end{figure}

Fig.~\ref{fig:total1_cyclic} shows the CSD and CSC graph plots of the noise data selected at random from Dataset-1 and the noise generated by each GAN-based model. Each GAN-based model learned the correlations of the cyclic frequencies at 122, 244, 366, 488, 610, and 732 Hz. However, the NGGAN performed better than the other GAN-based models.

Table~\ref{tab:statistic_1} presents a statistical analysis of the features of the noise samples generated by each GAN-based model. 
The NGGAN outperformed the other GAN-based models for seven of the nine performance metrics listed in Table~\ref{tab:statistic_1} (a) (mean). 
Both the NGGAN and DCGAN achieved the top spot in four of the nine performance metrics listed in Table~\ref{tab:statistic_1} (b) (standard deviation). 
The NGGAN achieved the best performance for seven of the nine performance metrics listed in Table~\ref{tab:statistic_1} (c). 
The NGGAN outperformed the DCGAN in terms of the maximum value (+9$\%$), energy value (+34$\%$), and standard deviation (+15$\%$).

Table~\ref{tab:cyclic_statistic_1} (a) presents the statistics of cyclic autocorrelation coefficients exceeding 0.9 of noise generated by the four GAN-based models (15,000 samples) for cyclic frequencies of 122, 244, 366, 488, 610, and 732 Hz over the frequency range of 0 to 200 kHz. 
The table lists the average number of autocorrelation coefficients that exceed 0.9. 
Because all of the cyclic autocorrelation coefficients in Dataset-1 exceed 0.9, a higher percentage of autocorrelation coefficients of the generated data exceeding 0.9 indicates a higher similarity with Dataset-1.
The error was calculated over the frequency range of 0-732 kHz. 
Statistical analysis showed that the NGGAN had the lowest accumulated error (33$\%$).

Table~\ref{tab:cyclic_statistic_1} (b) presents the distribution of the maximum autocorrelation coefficients.
If the number of maximum autocorrelation coefficients of the GAN-based models is closer to that of Dataset-1, it indicates a better learning performance in its cyclic spectral properties.
Because the maximum autocorrelation coefficients of all GAN-based models were located at a cyclic frequency of 122 Hz, we focused on the maximum autocorrelation coefficients located at 244, 366, 488, 610, and 732 Hz.
The NGGAN outperformed the other GAN-based models in four statistical quantities with errors of 5$\%$.

\begin{table}[tbh]
\caption{Dataset-1 Noise feature statistics: (a) mean, (b) standard deviation, and (c) median analysis.
Referring to Eqs. (1) to (9) features include (1) maximum samples [v], (2) mean [mv], (3) energy [mJ], (4) standard deviation [v], (5) skewness, (6) kurtosis, (7) count of samples with peak $>$ 0.05 V, (8) skewness of autocorrelation, and (9) kurtosis of autocorrelation.
}
\centering

\subfloat[][]{
\scalebox{0.97}{
\begin{tabular}{|c|c|c|c|c|c|} 
\hline
    \textbf{Feature} & \textbf{Dataset-1} &\textbf{~\cite{Letizia2020}} &\textbf{~\cite{Radford2015}} &
    \textbf{~\cite{Chien2021}} &\textbf{NGGAN} \\\hline
    \textbf{(1)} & 6.63E-1 & 7.48E-1 & 7.17E-1 & 5.23E-1& \textbf{6.46E-1}
     \\   \hline
    \textbf{(2)} & 2.14E-4 & $-2.74$E-3 & \textbf{4.81E-4} & 0.928 & 0.887
                                                \\  \hline
    \textbf{(3)} & 1.79 & 2.51 & 2.38 & 1.46 & \textbf{1.90}
                                                \\  \hline
    \textbf{(4)} & 4.23E-2 & 5.00E-2 & 4.75E-2 & 3.79E-2  & \textbf{4.36E-2}
                                                \\  \hline
    \textbf{(5)} & 4.42E-3 & $-$3.99E-3 & \textbf{6.80E-3} & 3.64E-1  & $-$2.22E-1
                                                \\  \hline
    \textbf{(6)} & 60.8 & 41.3  & 45.8 & 34.8 & \textbf{57.4}
                                                \\  \hline
    \textbf{(7)} & 266 & 630 & 512 & 399 & \textbf{317}
                                                \\  \hline
    \textbf{(8)} & 2.59  & 2.24 & 2.37 & 2.35 & \textbf{2.60}
                                                \\  \hline
    \textbf{(9)} & 10.9 & 9.32 & 9.89 & 9.94 & \textbf{10.9}
                                                \\  \hline
\end{tabular}}
}
\\
\subfloat[][]{
\scalebox{1}{
\begin{tabular}{|c|c|c|c|c|c|}  
\hline
    \textbf{Feature} & \textbf{Dataset-1} &\textbf{~\cite{Letizia2020}} &\textbf{~\cite{Radford2015}} &
    \textbf{~\cite{Chien2021}} & \textbf{NGGAN} 
                                               \\ \hline
    \textbf{(1)} & 7.59E-2 & 1.21E-1 & 2.03E-1 & 1.09E-1 & \textbf{6.98E-2}
                                                \\    \hline
    \textbf{(2)} & 1.20E-1 & 2.84E-1 & 2.78E-1 & \textbf{1.21E-1} & 2.95E-1                                                \\    \hline
    \textbf{(3)} & 3.87E-2 & \textbf{1.55E-1} & 1.12 & 3.97E-1  & 1.88E-1
                                                \\    \hline
    \textbf{(4)} & 4.56E-4 & \textbf{1.54E-3} & 1.08E-2 & 5.06E-3 & 2.14E-3
                                                \\    \hline
    \textbf{(5)} & 4.92E-1 & \textbf{4.59E-1} & 4.54E-1 & 2.83E-1 & 4.02E-1
                                                \\    \hline
    \textbf{(6)} & 4.52 & 6.72 & 5.81 & 5.88  & \textbf{3.79}
                                                \\    \hline
    \textbf{(7)} & 10.6 & \textbf{37.3} & 245 & 911 & 40.2
                                                \\    \hline
    \textbf{(8)} & 1.44E-1 & 1.03E-1 & 1.56E-1 & 3.58E-1 & \textbf{1.46E-1}
                                                \\    \hline
    \textbf{(9)} & 6.94E-1 & 4.89E-1 & 7.52E-1 & 1.40 & \textbf{7.09E-1}
                                                \\    \hline
\end{tabular}}
}
\\
\subfloat[][]{
\scalebox{0.95}{
\begin{tabular}{|c|c|c|c|c|c|}  
\hline
    \textbf{Feature} & \textbf{Dataset-1} &\textbf{~\cite{Letizia2020}} &\textbf{~\cite{Radford2015}} &
    \textbf{~\cite{Chien2021}} &\textbf{NGGAN} 
                                               \\\hline
    \textbf{(1)} & 6.54E-1 & 7.31E-1 & 6.88E-1 & 5.0E-1 & \textbf{6.40E-1}                                                \\    \hline
    \textbf{(2)} & $-$2.48E-4 & $-$5.81E-3 & \textbf{$-$1.19E-3} & 9.21E-1 & 8.61E-1                                   \\    \hline
    \textbf{(3)} & 1.79 & 2.50 & 2.14 & 1.41  & \textbf{1.90}
                                                \\    \hline
    \textbf{(4)} & 4.23E-2 & 5.00E-2 & 4.63E-2 & 3.75E-2 & \textbf{4.35E-2}
                                                \\    \hline
    \textbf{(5)} & 7.84E-4 & $-$5.33E-3 & \textbf{5.46E-3} & 3.61E-1 & $-$2.21E-1
                                                \\    \hline
    \textbf{(6)} & 60.4 & 40.2& 45.3& 34.1& \textbf{57.1}
                                                \\    \hline
    \textbf{(7)} & 266 & 629 & 452 & 387 & \textbf{314}
                                                \\    \hline
    \textbf{(8)} & 2.59 & 2.24 & 2.37 & 2.36 & \textbf{2.60}
                                                \\    \hline
    \textbf{(9)} & 10.9& 9.32 & 9.90 & 9.99 & \textbf{10.9}
                                                \\    \hline
\end{tabular}}
}
\label{tab:statistic_1}
\end{table}

\begin{table}[tbh]
\centering
\caption{Statistical comparisons for Dataset-1: (a) autocorrelation coefficients exceeding 0.9. (b) distribution of the maximum autocorrelation coefficient in the cyclic spectral at 122 Hz.} 
\subfloat[][]{
    \scalebox{1}{
    \begin{tabular}{|c|c|c|c|c|} 
    \hline
\textbf{Feature} & \textbf{~\cite{Letizia2020}} &
\textbf{~\cite{Radford2015}} &
\textbf{~\cite{Chien2021}} &
\textbf{NGGAN} \\ \hline
\textbf{122 Hz} & 94$\%$ & 99$\%$ & 96$\%$ & \textbf{100$\%$} \\ \hline
\textbf{244 Hz} & 79$\%$ & 93$\%$ & 84$\%$ & \textbf{96$\%$} \\ \hline
\textbf{366 Hz} & 54$\%$ & 57$\%$ & 55$\%$ & \textbf{94$\%$} \\ \hline
\textbf{488 Hz} & 59$\%$ & 43$\%$ & 49$\%$ & \textbf{93$\%$} \\ \hline
\textbf{610 Hz} & 71$\%$ & 41$\%$ & 58$\%$ & \textbf{92$\%$} \\ \hline
\textbf{732 Hz} & 91$\%$ & 41$\%$ & 74$\%$ & \textbf{92$\%$} \\ \hline
\textbf{Error} & 152$\%$ & 226$\%$ & 184$\%$ & \textbf{33$\%$} \\ \hline 
    \end{tabular}}
}
\\
\subfloat[][]{
    \scalebox{1}{
    \begin{tabular}{|c|c|c|c|c|c|} 
     \hline
    \textbf{Feature} & \textbf{Dataset-1} & \textbf{~\cite{Letizia2020}} &
\textbf{~\cite{Radford2015}} &
\textbf{~\cite{Chien2021}} &
\textbf{NGGAN} \\ \hline  
    \textbf{0-50 kHz} & 42$\%$ & 22$\%$ & 40$\%$ & 26$\%$ & \textbf{42$\%$} \\ \hline
    \textbf{50-100 kHz} & 1$\%$ & \textbf{0$\%$} & 2$\%$ & 5$\%$ & 2$\%$ \\ \hline
    \textbf{100-150 kHz} & 6$\%$ & 69$\%$ & 9$\%$ & 11$\%$ & \textbf{7$\%$} \\ \hline
    \textbf{150-200 kHz} & 51$\%$ & 9$\%$ & \textbf{48$\%$} & 57$\%$ & \textbf{48$\%$} \\ \hline
    \textbf{Error} & - & 126$\%$ & 9$\%$ & 31$\%$ & \textbf{5$\%$} \\ \hline    
    \end{tabular}}
}    
\label{tab:cyclic_statistic_1}
\end{table}

\begin{figure}[tbh]
  \centering
  \includegraphics[width=.45\textwidth]{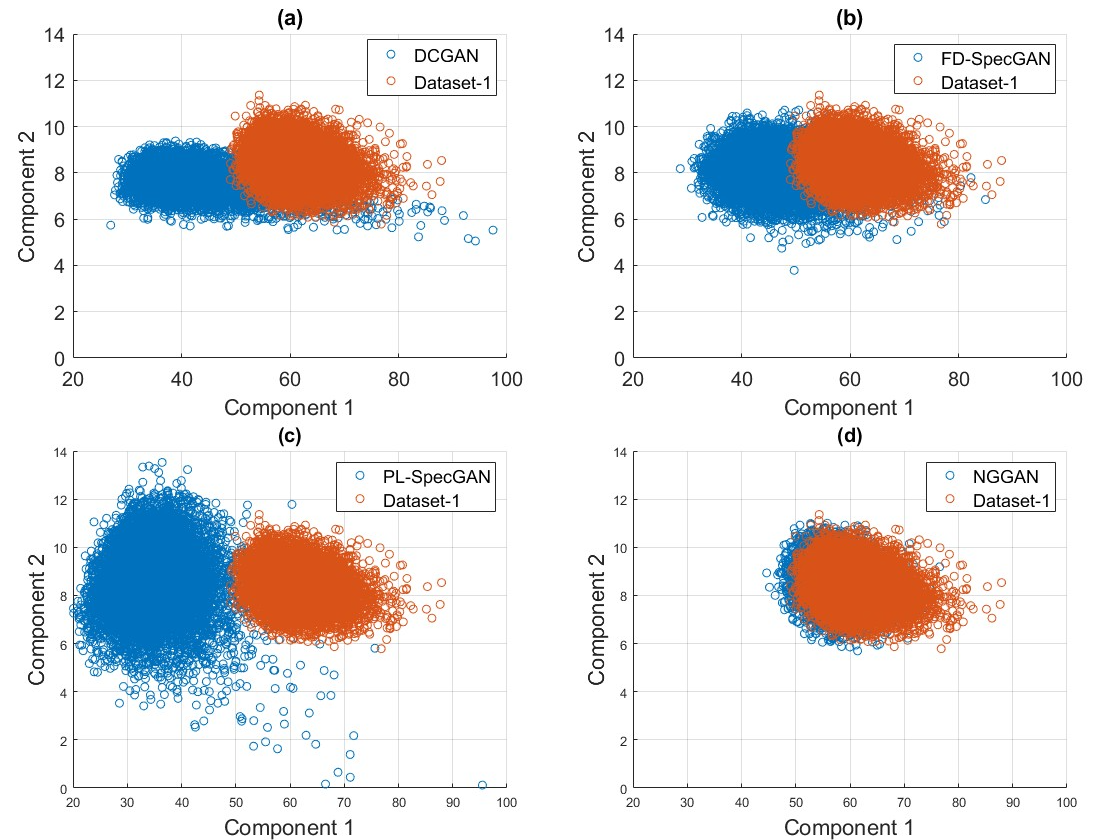}
  \caption{PCA scatter: (a) DCGAN~\cite{Letizia2020}; (b) FD-SpecGAN~\cite{Radford2015}; (c) PL-SpecGAN~\cite{Chien2021}; (d) NGGAN.}
  \label{fig:pca_1}
\end{figure}

Fig.~\ref{fig:pca_1} presents the PCA scatter plots of the noise generated by Dataset-1 and each GAN-based model. 
The X and Y-axes correspond to the first and second principal components, respectively.
Thus, the NGGAN was the most effective at generating noise for learning Dataset-1. 
The FID values in Table~\ref{tab:FID_1} indicate that the NGGAN achieved the best balance between the quality and diversity of the generated noise.

\begin{table}[tbh]
\centering
\caption{PCA feature FID analysis for Dataset-1.}

    \scalebox{1}{
    \begin{tabular}{|c|c|c|c|} 
    \hline

    \textbf{DCGAN~\cite{Letizia2020}} &\textbf{FD-SpecGAN~\cite{Radford2015}} &\textbf{PL-SpecGAN~\cite{Chien2021}} &\textbf{NGGAN}
    \\
    \hline
     387.09 & 225.48 & 672.53  & \textbf{12.16}
    \\
    \hline
    \end{tabular}}
    
\label{tab:FID_1}
\end{table}

\subsection{Dataset-2 (FRESH-Generated)}

\begin{figure}[tbh]
\centering
  \includegraphics[width=0.45\textwidth]{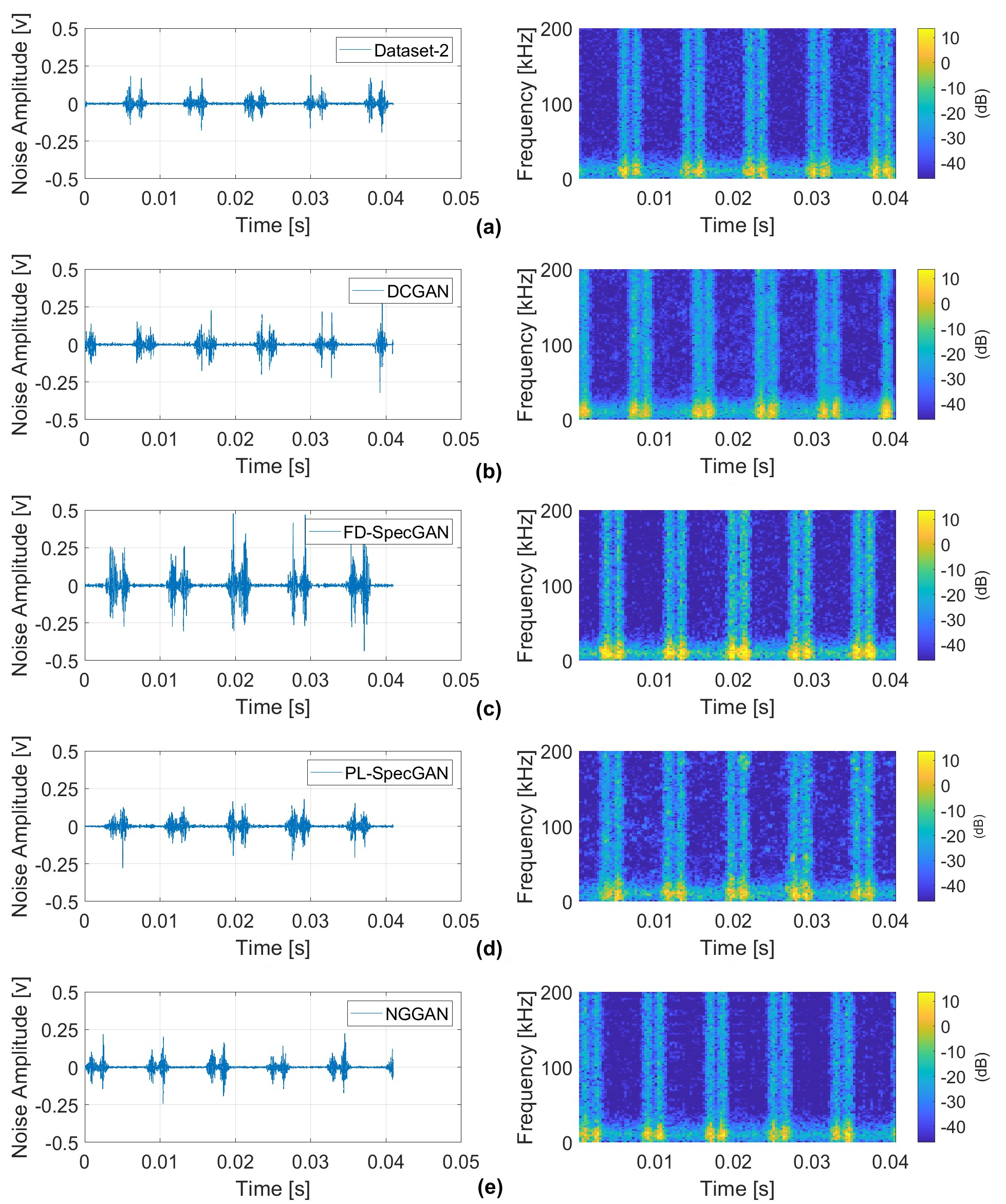}
  \caption{Noise time series and spectrograms: (a) Dataset-2; (b) DCGAN~\cite{Letizia2020}; (c) FD-SpecGAN~\cite{Radford2015}; (d) PL-SpecGAN~\cite{Chien2021}; (e) NGGAN.}
  \label{fig:trace_spectrogram_2}
\end{figure}

Fig.~\ref{fig:trace_spectrogram_2} presents the noise time series and corresponding spectrograms for the four GAN-based models learning Dataset-2.
As in the time series and spectrogram analyses in the previous sub-section (Dataset-1), the frequency components and pulse patterns of the generated noise varied cyclically with time. 
The FD-SpecGAN model generated larger pulse values with wide waveform variations and notable statistical inconsistencies. 
The PL-SpecGAN-generated noise time series was significantly stable in terms of the pulse amplitude and trajectory.

Table~\ref{tab:statistic_2} presents a statistical analysis of the features of the noise samples generated by each GAN-based model.
The NGGAN achieved the best performance for eight of the nine performance metrics listed in Table~\ref{tab:statistic_2} (a) (mean). 
The NGGAN achieved the best performance for eight of the nine performance metrics listed in Table~\ref{tab:statistic_2} (b) (standard deviation).
The NGGAN outperformed the DCGAN in terms of the maximum value (+37$\%$), energy value (+41$\%$), and standard deviation (+17$\%$). 
Because of the phase loss from the Griffin-Lim process, the FD-SpecGAN-generated noise is the most unstable, resulting in errors in the maximum and energy values exceeding those of other GAN-based models. 
In contrast, the PL-SpecGAN is not affected by the phase loss of the Griffin-Lim process, which has better results than the FD-SpecGAN.

\begin{table}[tbh]
\caption{Dataset-2 Noise feature statistics: (a) mean, (b) standard deviation and (c) median analysis.
Referring to Eqs. (1) to (9) features include (1) maximum samples [v], (2) mean [mv], (3) energy [mJ], (4) standard deviation [v], (5) skewness, (6) kurtosis, (7) count of samples with peak $>$ 0.05 V, (8) skewness of autocorrelation, and (9) kurtosis of autocorrelation.
}
\centering

\subfloat[][]{
\scalebox{1}{
\begin{tabular}{|c|c|c|c|c|c|} 
\hline
    \textbf{Feature} & \textbf{Dataset-2} &\textbf{~\cite{Letizia2020}} &\textbf{~\cite{Radford2015}} &
    \textbf{~\cite{Chien2021}} &\textbf{NGGAN} 
                                               \\\hline

    \textbf{(1)} & 2.07E-1 & 2.86E-1 & 7.15E-1 & 1.72E-1& \textbf{2.05E-1}
                                                \\
    \hline
    \textbf{(2)} & $-8.98$E-3 & 3.19E-4 & \textbf{-2.39E-3} & 4.40E-1 & $-1.29$E-1
                                                \\
    \hline
    \textbf{(3)} & 5.77E-1 & 8.28E-1 & 3.13E1 & 4.47E-1 & \textbf{5.67E-1}
                                                \\
    \hline
    \textbf{(4)} & 2.40E-2 & 2.87E-2 & 8.74E-2 & 2.06E-2  & \textbf{2.37E-2}
                                                \\
    \hline
    \textbf{(5)} & $-3.35$E-2 & 1.59E-3 & 2.93E-3 & 1.41E-1  & \textbf{9.82E-3}
                                                \\
    \hline
    \textbf{(6)} & 1.76E1 & 2.28E1  & 1.49E1 & 1.47E1 & \textbf{1.74E1}
                                                \\
    \hline
    \textbf{(7)} & 1.73E2 & 2.07E2 & 1.94E2 & 1.54E2 & \textbf{1.68E2}
                                                \\
    \hline
    \textbf{(8)} & 2.59E-1  & 4.32E-1 & 5.12E-1 & 3.72E-1 & \textbf{2.66E-1}
                                                \\
    \hline
    \textbf{(9)} & 1.55 & 1.63 & 2.31 & 1.62 & \textbf{1.55}
                                                \\
    \hline
\end{tabular}}
}
\\
\subfloat[][]{
\scalebox{1}{
\begin{tabular}{|c|c|c|c|c|c|}  
\hline
    \textbf{Feature} & \textbf{Dataset-2} &\textbf{~\cite{Letizia2020}} &\textbf{~\cite{Radford2015}} &
    \textbf{~\cite{Chien2021}} & \textbf{NGGAN} 
                                               \\
\hline
    \textbf{(1)} & 3.10E-2 & 6.63E-2 & 1.22 & 4.99E-2 & \textbf{3.71E-2}  \\ \hline
    \textbf{(2)} & 3.64E-1 & 1.81E-1 & 4.95E-1 & 1.74E-1 & \textbf{2.95E-1}  \\  \hline
    \textbf{(3)} & 7.99E-2 & 1.58E-1 & 1.13E2 & 2.03E-1  & \textbf{9.37E-2} \\  \hline
    \textbf{(4)} & 1.66E-3 & 2.56E-3 & 1.54E-1 & 4.48E-3 & \textbf{2.00E-3}    \\  \hline
    \textbf{(5)} & 4.31E-1 & 6.59E-1 & 2.93E-1 & 2.91E-1 & \textbf{4.18E-1}  \\    \hline
    \textbf{(6)} & 2.82 & 7.06 & 4.86 & 3.11  & \textbf{2.90}  \\   \hline
    \textbf{(7)} & 2.17E1 & \textbf{2.4E1} & 1.71E2 &  5.66E1 & 42.61E1 \\   \hline
    \textbf{(8)} & 5.15E-2 & 6.34E-2 & 6.18E-1 & 7.55E-2 & \textbf{5.33E-2} \\  \hline
    \textbf{(9)} & 1.73E-2 & 4.20E-2 & 2.20 & 5.87E-2 & \textbf{1.82E-2} \\  \hline
\end{tabular}}
}
\\
\subfloat[][]{
\scalebox{1}{
\begin{tabular}{|c|c|c|c|c|c|}  
\hline
    \textbf{Feature} & \textbf{Dataset-2} &\textbf{~\cite{Letizia2020}} &\textbf{~\cite{Radford2015}} &
    \textbf{~\cite{Chien2021}} &\textbf{NGGAN} 
                                               \\
\hline
    \textbf{(1)} & 2.05E-1 & 2.75E-1 & \textbf{2.08E-1} & 1.65E-1 & 22.01E-1 \\  \hline
    \textbf{(2)} & $-8.75$E-3 & 2.14E-5 & 55.57E-5 & 4.17E-1 & $-1.56$E-1 \\ \hline
    \textbf{(3)} & 5.73E-1 & 8.06E-1 & 6.06E-1 & 4.06E-1 & \textbf{5.65E-1} \\ \hline
    \textbf{(4)} & 2.39E-2 & 2.84E-2 & 2.46E-2 & 2.01E-2 & \textbf{2.38E-2} \\ \hline
    \textbf{(5)} & $-1.86$E-2 & 2.32E-3 & \textbf{1.27E-3} & 1.35E-1 & 3.06E-3 \\ \hline
    \textbf{(6)} & 1.72E1 & 2.12E1& 1.53E1& 1.43E1& \textbf{1.69E1} \\ \hline
    \textbf{(7)} & 1.72E2 & 2.05E2 & 1.84E2 & 1.51E2 & \textbf{1.69E2} \\ \hline
    \textbf{(8)} & 2.57E-1 & 4.33E-1 & 2.95E-1 & 3.70E-1 & \textbf{2.65E-1} \\ \hline
    \textbf{(9)} & 1.54& 1.62 & 1.56 & 1.61 & \textbf{1.55} \\ \hline
\end{tabular}}
}
\label{tab:statistic_2}
\end{table}

In Dataset-2, the pulse noise period ${T_{ac}}/{2}$ was 1/122 s, indicating cycling frequencies of 122, 244, 366, 488, 610, and 732 Hz.
According to~\cite{Nieman2013}, the correlation coefficient of practical NB-PLC noise decreases inversely with cycling frequency. 
However, the correlation coefficients of Dataset-2 noise deviate significantly from those of the practical NB-PLC noise measurements.
At 244 and 366 Hz, the correlation coefficients of the Dataset-2 noise were relatively low compared to practical NB-PLC noise.
At 488 Hz, the correlation coefficients of Dataset-2 noise were relatively high with practical NB-PLC noise.

\begin{figure}[tbh]
  \centering
  \includegraphics[width=.45\textwidth]{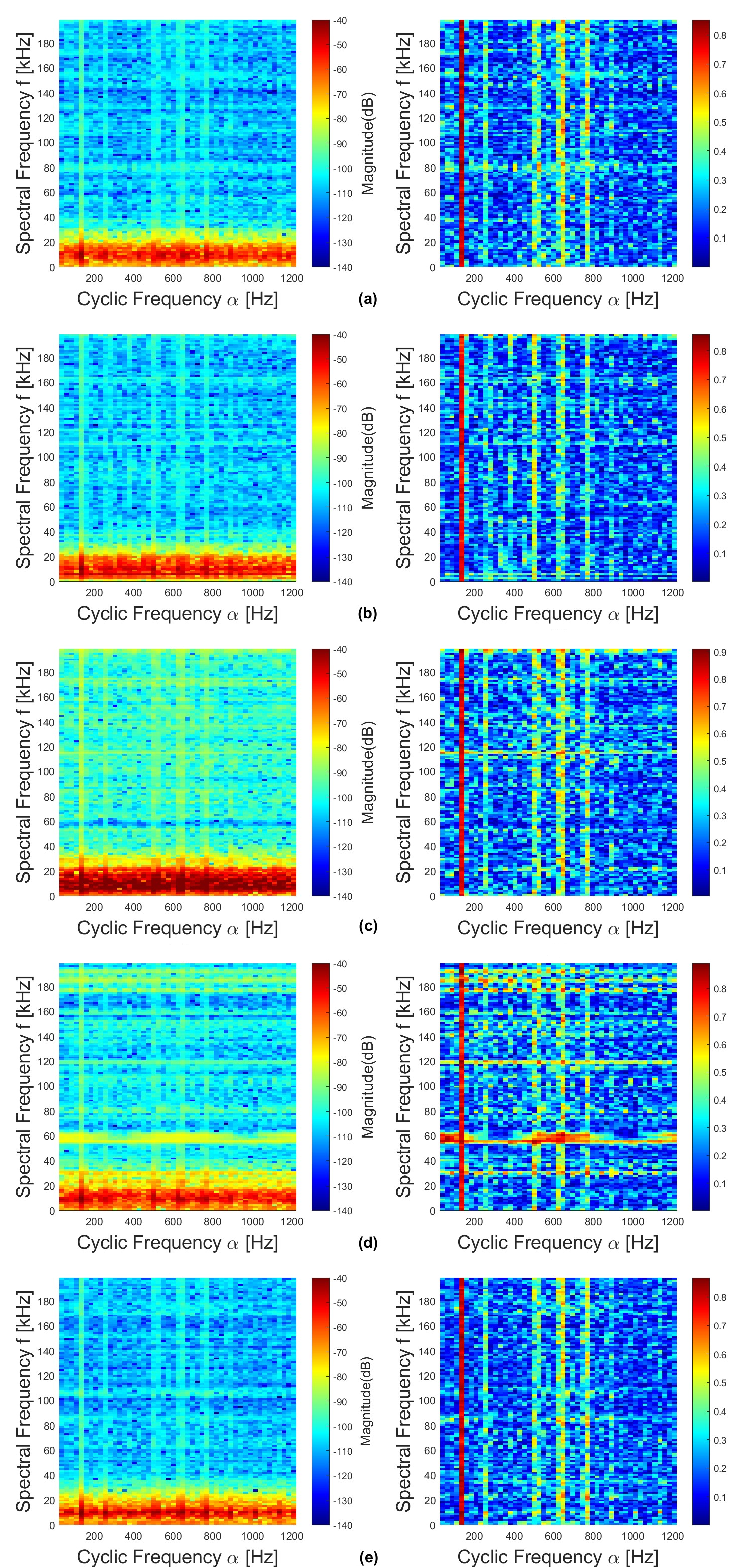}
  \caption{CSD and CSC graphs: (a) Dataset-2; (b) DCGAN~\cite{Letizia2020}; (c) FD-SpecGAN~\cite{Radford2015}; (d) PL-SpecGAN~\cite{Chien2021}; (e) NGGAN.}
  \label{fig:total2_cyclic}
\end{figure}

Fig.~\ref{fig:total2_cyclic} presents the CSD and CSC graph plots of the noise data selected at random from Dataset-1 and the noise generated by each GAN-based model.
The CSD and CSC graphs demonstrate that all the GAN-based models learned well patterns associated with a cycling frequency of 122 Hz.
However, the learning performance of the GAN-based models varied considerably at other cycling frequencies.

Table~\ref{tab:cyclic_statistic_2} (a) presents the statistics of cyclic autocorrelation coefficients exceeding 0.5 of noise generated by the four GAN-based models (15,000 samples) for the cyclic frequencies of 122, 244, 366, 488, 610, and 732 Hz.
The table lists the average number of autocorrelation coefficients that exceed 0.5.
By calculating the average number of correlation coefficients exceeding 0.5 over a frequency range of 0-200 kHz, we determined that the most important features were those associated with a cycling frequency of 122 Hz.
The error was calculated over the frequency range of 0-122 kHz.
Similarly, the four GAN-based models performed well in extracting features, as indicated by the error results.

Table~\ref{tab:cyclic_statistic_2} (b) presents the distributions of the maximum autocorrelation coefficients. 
For all the GAN-based models, the maximum autocorrelation coefficients were located at a cyclic frequency of 122 Hz.
Therefore, we focused on the maximum autocorrelation coefficients at 244 Hz, 366 Hz, 488 Hz, 610 Hz, and 732 Hz.
It is clear that the FD-SpecGAN outperformed all other GAN-based models, as indicated by a cumulative error of 7$\%$.

\begin{table}[tbh]
\centering
\caption{Statistical comparisons for Dataset-2: (a) autocorrelation coefficients exceeding 0.5. (b) distribution of the maximum autocorrelation coefficient in the cyclic spectral at 122 Hz.}
\subfloat[][]{
    \scalebox{1}{
    \begin{tabular}{|c|c|c|c|c|} 
    \hline
\textbf{Feature} & \textbf{~\cite{Letizia2020}} &
\textbf{~\cite{Radford2015}} &
\textbf{~\cite{Chien2021}} &
\textbf{NGGAN} \\ \hline
\textbf{122 Hz} & \textbf{100$\%$} & 96$\%$ & \textbf{100$\%$} & \textbf{100$\%$} \\ \hline
\textbf{244 Hz} & \textbf{112$\%$} & 138$\%$ & 190$\%$ & 174$\%$ \\ \hline
\textbf{366 Hz} & 153$\%$ & 278$\%$ & 509$\%$ & \textbf{122$\%$} \\ \hline
\textbf{488 Hz} & \textbf{100$\%$} & 98$\%$ & 115$\%$ & 84$\%$ \\ \hline
\textbf{610 Hz} & 88$\%$ & 121$\%$ & 114$\%$ & \textbf{106$\%$} \\ \hline
\textbf{732 Hz} & \textbf{111$\%$} & 266$\%$ & 289$\%$ & 139$\%$ \\ \hline
\textbf{Error} & \textbf{0$\%$} & 4$\%$ & \textbf{0$\%$} & \textbf{0$\%$} \\ \hline 
    \end{tabular}}
}
\\
\subfloat[][]{
    \scalebox{1}{
    \begin{tabular}{|c|c|c|c|c|c|} 
     \hline
    \textbf{Feature} & \textbf{Dataset-2} & \textbf{~\cite{Letizia2020}} &
\textbf{~\cite{Radford2015}} &
\textbf{~\cite{Chien2021}} &
\textbf{NGGAN} \\ \hline  
    \textbf{0-20 kHz} & 14$\%$ & 9$\%$ & 10$\%$ & 2$\%$ & \textbf{12$\%$} \\ \hline
    \textbf{20-80 kHz} & 30$\%$ & 22$\%$ & \textbf{30$\%$} & \textbf{30$\%$} & 20$\%$ \\ \hline
    \textbf{80-140 kHz} & 28$\%$ & 33$\%$ & \textbf{29$\%$} & 35$\%$ & 35$\%$ \\ \hline
    \textbf{140-200 kHz} & 29$\%$ & 36$\%$ & \textbf{31$\%$} & 33$\%$ & 32$\%$ \\ \hline
    \textbf{Error} & - & 25$\%$ & \textbf{7$\%$} & 23$\%$ & 22$\%$ \\ \hline    
    \end{tabular}}
}    
\label{tab:cyclic_statistic_2}
\end{table}

\begin{figure}[tbh]
  \centering
  \includegraphics[width=.45\textwidth]{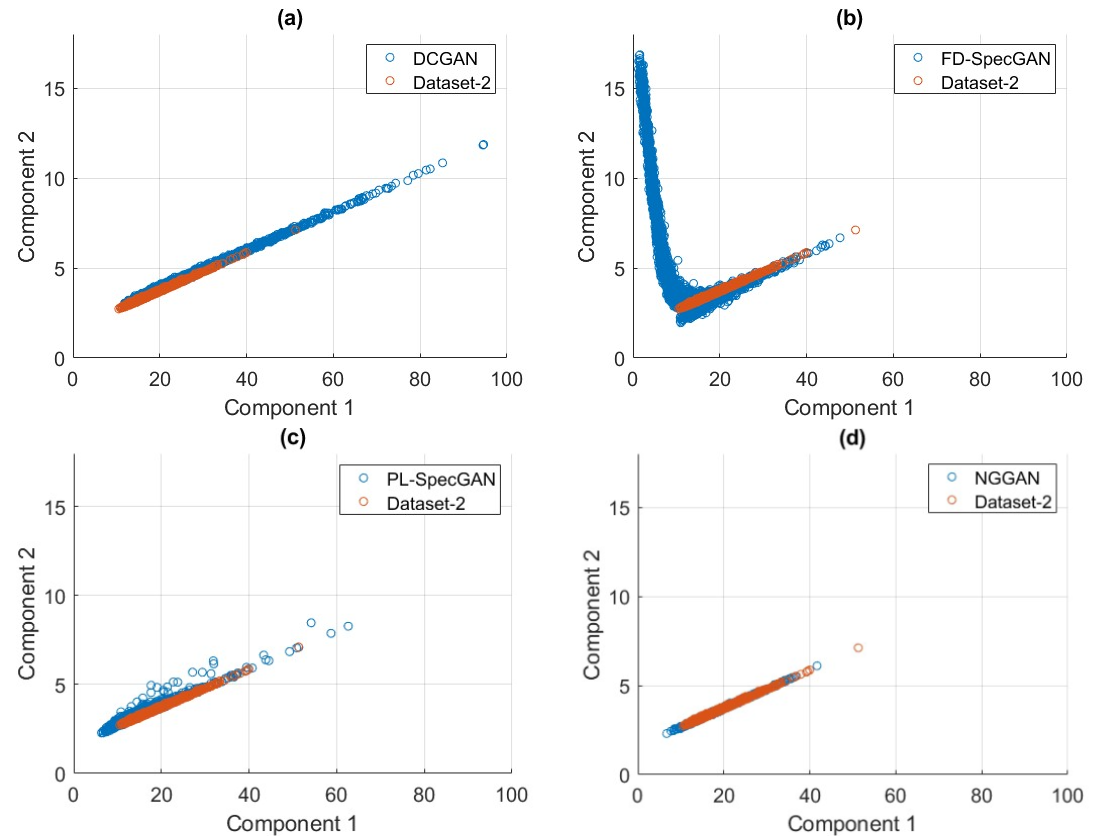}
  \caption{PCA scatter: (a) DCGAN~\cite{Letizia2020}; (b) FD-SpecGAN~\cite{Radford2015}; (c) PL-SpecGAN~\cite{Chien2021}; (d) NGGAN.}
  \label{fig:pca_2}
\end{figure}

Fig.~\ref{fig:pca_2} shows the PCA scatter plots of the noise generated by Dataset-2 and each GAN-based model.
The NGGAN-generated noise and PL-SpecGAN-generated noise exhibited the greatest similarity with Dataset-2.
The FID values in Table~\ref{tab:FID_2} show that the NGGAN provides a suitable balance between the quality and diversity of generated noise.
The FID value of the PL-SpecGAN-generated noise was superior to that of the FD-SpecGAN.

\begin{table}[tbh]
\centering
\caption{PCA feature FID analysis for Dataset-2.}

    \scalebox{1}{
    \begin{tabular}{|c|c|c|c|} 
    \hline

    \textbf{DCGAN~\cite{Letizia2020}} &\textbf{FD-SpecGAN~\cite{Radford2015}} &\textbf{PL-SpecGAN~\cite{Chien2021}} &\textbf{NGGAN}
    \\
    \hline
     45.15 & 18.66 & 8.48  & \textbf{0.07}
    \\
    \hline
    \end{tabular}}
\label{tab:FID_2}
\end{table}

\subsection{Dataset-3 (Experimentally Measured Data)}

\begin{figure}[tb]
\centering
  \includegraphics[width=0.45\textwidth]{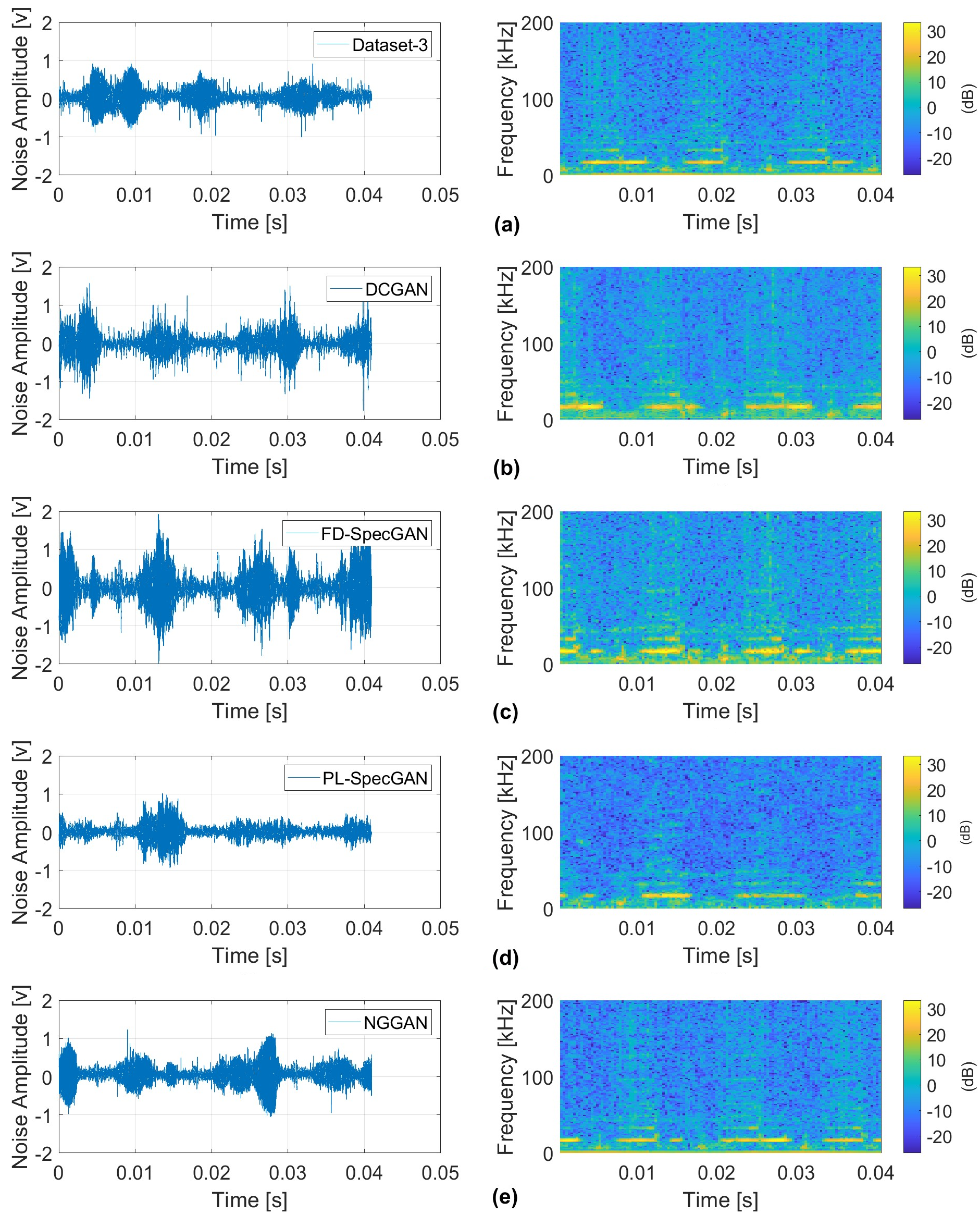}
  \caption{Noise time series and spectrograms: (a) Dataset-3; (b) DCGAN~\cite{Letizia2020}; (c) FD-SpecGAN~\cite{Radford2015}; (d) PL-SpecGAN~\cite{Chien2021}; (e) NGGAN.}
  \label{fig:trace_spectrogram_3}
\end{figure}

Fig.~\ref{fig:trace_spectrogram_3} presents the noise time series and the corresponding spectrograms for the four GAN-based models learning Dataset-3.
The spectrograms revealed complex temporal traces with noise impulses occurring in bursts at intervals of roughly every 8.3 ms over a frequency range of up to 75 kHz. 
In addition, the frequency components vary with time.
However, periodic variations occur within specific frequency bands. 
The time series revealed that the burst impulses varied randomly with time.
The complexity of these trace patterns makes it difficult to observe their structural features. 
The NGGAN-generated and PL-SpecGAN-generated noises showed the greatest similarities for Dataset-3 learning.

Table~\ref{tab:statistic_3} presents the statistical analysis of the noise features generated by each GAN-based model. 
The NGGAN achieved the best performance in five of the nine performance metrics listed in Table~\ref{tab:statistic_3} (a) (mean), whereas the FD-SpecGAN achieved the best performance in three of the nine performance metrics. 
The NGGAN achieved the best performance in seven of the nine performance metrics listed in Tables~\ref{tab:statistic_3} (b) (standard deviation) and~\ref{tab:statistic_3} (c) (median). 
The NGGAN outperformed the DCGAN in terms of the maximum value (+6$\%$), energy value (+80$\%$), and standard deviation (+5$\%$). 
Additionally, the FD-SpecGAN performed better on Dataset-3 compared with Dataset-2.
There were no significant differences between the PL-SpecGAN and FD-SpecGAN.

\begin{table}[tbh]
\caption{Dataset-3 Noise feature statistics: (a) mean, (b) standard deviation and (c) median analysis.
Referring to Eqs. (1) to (9) features include (1) maximum samples [v], (2) mean [mv], (3) energy [mJ], (4) standard deviation [v], (5) skewness, (6) kurtosis, (7) count of samples with peak $>$ 0.05 V, (8) skewness of autocorrelation, and (9) kurtosis of autocorrelation.
}
\centering

\subfloat[][]{
\scalebox{0.92}{
\begin{tabular}{|c|c|c|c|c|c|} 
\hline
    \textbf{Feature} & \textbf{Dataset-3} &\textbf{~\cite{Letizia2020}} &\textbf{~\cite{Radford2015}} &
    \textbf{~\cite{Chien2021}} &\textbf{NGGAN} \\ \hline

    \textbf{(1)} & 3.33 & 3.96 & 2.01 & 2.16& \textbf{2.95} \\  \hline
    \textbf{(2)} & 2.15E2 & $-1.37$E-2 & $-9.64$E-3 & 6.84E1 & \textbf{1.77E-2}  \\ \hline
    \textbf{(3)} & 2.63E2 & 4.88E2 & 2.50E2 & 1.87E2 & \textbf{2.52E2}  \\  \hline
    \textbf{(4)} & 4.62E-1 & 6.96E-1 & \textbf{4.61E-1} & 4.22E-1  & 4.56E-1  \\   \hline
    \textbf{(5)} & 3.21E-2 & $-5.39$E-4 & 5.74E-4 & \textbf{3.71E-2}  & $-1.84$E-3  \\  \hline
    \textbf{(6)} & 6.08E1 & 4.13E1  & 4.58E1 & 3.48E1 & \textbf{5.74E1} \\ \hline
    \textbf{(7)} & 2.66E2 & 6.30E2 & 5.12E2 & 3.99E2 & \textbf{3.17E2} \\ \hline
    \textbf{(8)} & 2.59  & 2.24 & 2.37 & 2.35 & \textbf{2.60}   \\ \hline
    \textbf{(9)} & 1.09E1 & 9.32 & 9.89 & 9.94 & \textbf{1.09E1} \\
    \hline
\end{tabular}}
}
\\
\subfloat[][]{
\scalebox{1}{
\begin{tabular}{|c|c|c|c|c|c|}  
\hline
    \textbf{Feature} & \textbf{Dataset-3} &\textbf{~\cite{Letizia2020}} &\textbf{~\cite{Radford2015}} &
    \textbf{~\cite{Chien2021}} & \textbf{NGGAN} 
                                               \\
\hline
    \textbf{(1)} & 4.46E-1 & 5.18E-1 & 7.91E-1 & 3.98E-1 & \textbf{4.11E-1}  \\ \hline
    \textbf{(2)} & 4.22E1 & 3.26 & 1.81 & 8.84 & \textbf{3.03E1}  \\  \hline
    \textbf{(3)} & 6.42E1 & 9.11E1 & 2.30E2 & 5.29E1 & \textbf{6.57E1} \\  \hline
    \textbf{(4)} & 4.59E-2 & 6.15E-2 & 1.93E-1 & 6.05E-2 & \textbf{5.75E-2}    \\  \hline
    \textbf{(5)} & 6.72E-2 & 8.99E-2 & 5.23E-2 & 5.61E-2 & \textbf{6.33E-2}  \\    \hline
    \textbf{(6)} & 7.34E-1 & 5.00E-1 & 4.59E-1 & 8.95E-1  & \textbf{6.30E-1}  \\   \hline
    \textbf{(7)} & 2.34E2 & 1.28E3 & 1.43E2 &  2.02E2 & \textbf{2.24E2} \\   \hline
    \textbf{(8)} & 1.12E-1 & 8.26E-2 & \textbf{8.36E-2} & 6.30E-2 & 6.74E-2 \\  \hline
    \textbf{(9)} & 2.13E-1 & 2.60E-1 & \textbf{2.12E-1} & 1.65E-1 & 2.49E-1 \\  \hline
\end{tabular}}
}
\\
\subfloat[][]{
\scalebox{0.92}{
\begin{tabular}{|c|c|c|c|c|c|}  
\hline
    \textbf{Feature} & \textbf{Dataset-3} &\textbf{~\cite{Letizia2020}} &\textbf{~\cite{Radford2015}} &
    \textbf{~\cite{Chien2021}} &\textbf{NGGAN} 
                                               \\
\hline
    \textbf{(1)} & 3.32 & 3.90 & 1.84 & 2.11 & \textbf{2.92} \\  \hline
    \textbf{(2)} & 2.04E2 & $-5.43$E-3 & $-7.87$E-3 & 6.78E1 & \textbf{1.67E2} \\ \hline
    \textbf{(3)} & 2.46E2 & 4.75E2 & 1.74E2 & 1.81E2 & \textbf{2.30E2} \\ \hline
    \textbf{(4)} & 4.53E-1 & 6.89E-1 & 4.17E-1 & 4.20E-1 & \textbf{4.49E-1} \\ \hline
    \textbf{(5)} & 3.24E-2 & $-1.53$E-3 & 2.05E-4 & \textbf{3.40E-2} & $-1.53$E-4 \\ \hline
    \textbf{(6)} & 4.47 & 4.74 & 4.10 & 4.90 & \textbf{4.23} \\ \hline
    \textbf{(7)} & 2.69E3 & 2.49E3 & \textbf{2.58E3} & 3.01E3 & 3.11E3 \\ \hline
    \textbf{(8)} & $-1.45$E-1 & $-1.21$E-1 & $-1.40$E-1 & \textbf{-1.47E-1} & \textbf{-1.43E-1} \\ \hline
    \textbf{(9)} & 1.66& 1.67 & 1.65 & 1.68 & \textbf{1.66} \\ \hline
\end{tabular}}
}
\label{tab:statistic_3}
\end{table}

\begin{figure}[tbh]
  \centering
  \includegraphics[width=.45\textwidth]{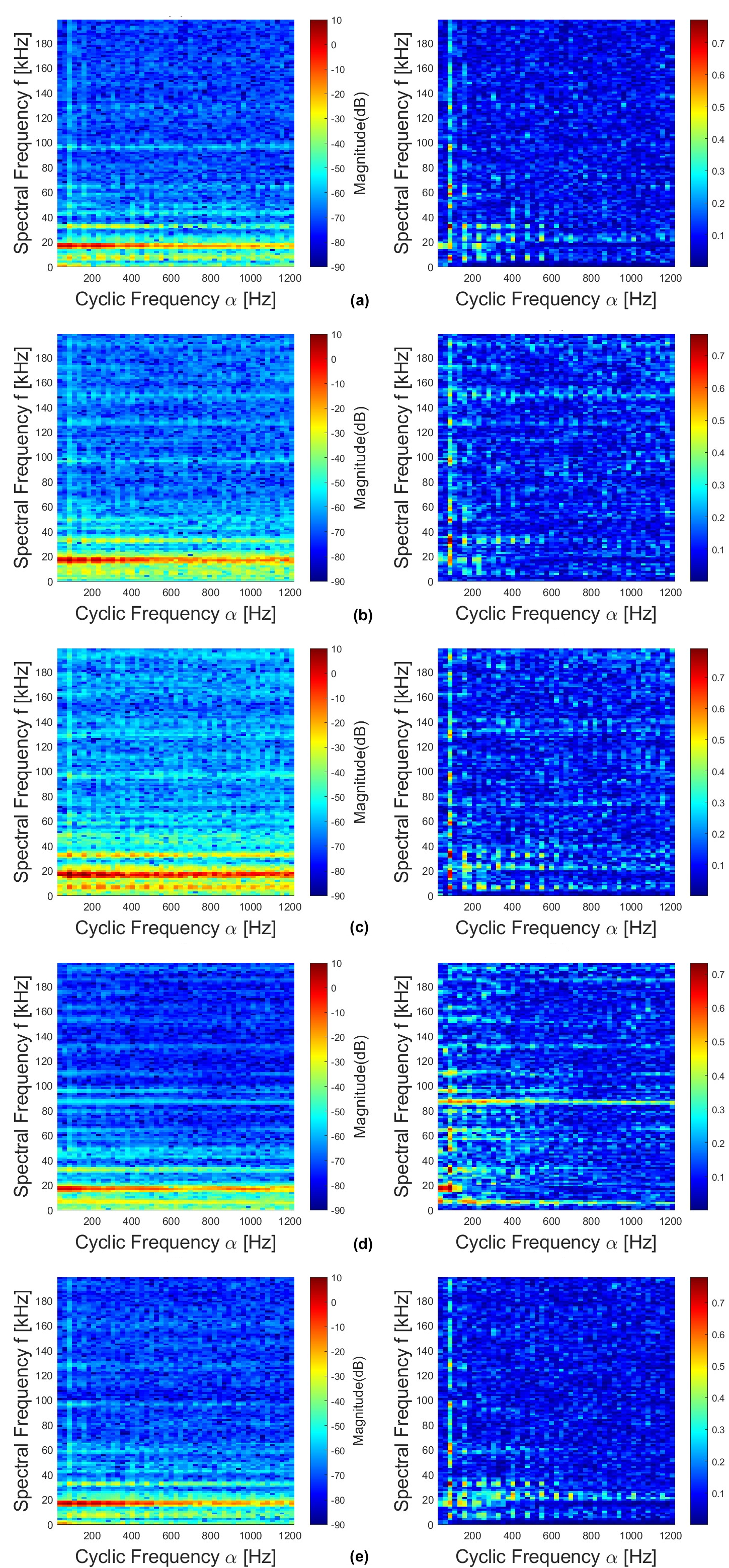}
  \caption{CSD and CSC graphs: (a) Dataset-3; (b) DCGAN~\cite{Letizia2020}; (c) FD-SpecGAN~\cite{Radford2015}; (d) PL-SpecGAN~\cite{Chien2021}; (e) NGGAN.}
  \label{fig:total3_cyclic}
\end{figure}

Fig.~\ref{fig:total3_cyclic} shows the CSD and CSC plots of the noise generated by each GAN-based model and noise data sampled randomly from Dataset-3.
In contrast to the mathematical models used in Dataset-1 and Dataset-2, the cyclo-stationary frequency of Dataset-3 was approximately 114 Hz.
The four GAN-based models learned correlation coefficients at 114 Hz, 228 Hz, 342 Hz, 456 Hz, 570 Hz, and 684 Hz.
However, the learning performances differed significantly at different cyclic frequencies.

Table~\ref{tab:cyclic_statistic_3} (a) presents the statistics of the autocorrelation coefficients exceeding 0.3 of noise generated by the four GAN-based models generated by the four GAN-based models (15,000 samples) for cycling frequencies of 114, 228, 342, 456, 570, and 684 Hz.
The error was calculated over the frequency range of 0-342 kHz. 
The table lists the average number of autocorrelation coefficients that exceed 0.3. 
Because all the cyclic autocorrelation coefficients in Dataset-3 exceed 0.3, the higher percentage of autocorrelation coefficients of the generated data exceeding 0.3 indicates a higher similarity with Dataset-3. 
Statistical analysis showed that the NGGAN had the lowest accumulated error (33$\%$).
The simulation results revealed that the most important features in Dataset-3 were associated with cyclo-stationary frequencies of 114, 228, and 342 Hz. 
The FD-SpecGAN achieved the best performance with a cumulative error of 31$\%$, whereas the NGGAN achieved similar performance.

Table~\ref{tab:cyclic_statistic_3} (b) presents the distribution of maximum autocorrelation coefficients. 
The maximum autocorrelation coefficients of the four GAN-based models were located at a cyclic frequency of 114 Hz.
Therefore, we focused on the maximum autocorrelation coefficients at 228, 342, 456, 570, and 684 Hz.
The FD-SpecGAN outperformed the other GAN-based models in learning Dataset-3.
However, the NGGAN achieved a performance similar to that of the FD-SpecGAN.
The PL-SpecGAN had the poorest performance. 
As shown in Tables~\ref{tab:cyclic_statistic_3} (a) and (b), the NGGAN outperformed the DCGAN by 37$\%$ and 5$\%$, respectively.  

\begin{table}[tbh]
\centering
\caption{Statistical comparisons for Dataset-3: (a) cyclic autocorrelation coefficients exceeding 0.3.
(b) distribution of maximum values of cyclic autocorrelation coefficients in cyclic spectral at 114 Hz.} 
\subfloat[][]{
    \scalebox{1}{
    \begin{tabular}{|c|c|c|c|c|} 
    \hline
\textbf{Feature} & \textbf{~\cite{Letizia2020}} &
\textbf{~\cite{Radford2015}} &
\textbf{~\cite{Chien2021}} &
\textbf{NGGAN} \\ \hline
\textbf{114 Hz} & 135$\%$ & \textbf{99$\%$} & 121$\%$ & 105$\%$ \\ \hline
\textbf{228 Hz} & 83$\%$ & 89$\%$ & 177$\%$ & \textbf{92$\%$} \\ \hline
\textbf{342 Hz} & \textbf{82$\%$} & 81$\%$ & 195$\%$ & 80$\%$ \\ \hline
\textbf{456 Hz} & 188$\%$ & 97$\%$ & 455$\%$ & \textbf{102$\%$} \\ \hline
\textbf{570 Hz} & 195$\%$ & 97$\%$ & 455$\%$ & \textbf{102$\%$} \\ \hline
\textbf{684 Hz} & 169$\%$ & \textbf{92$\%$} & 374$\%$ & 193$\%$ \\ \hline
\textbf{Error} & 70$\%$ & \textbf{31$\%$} & 193$\%$ & 33$\%$ \\ \hline 
\end{tabular}}
}
\\
\subfloat[][]{
    \scalebox{1}{
    \begin{tabular}{|c|c|c|c|c|c|} 
     \hline
    \textbf{Feature} & \textbf{Dataset-3} & \textbf{~\cite{Letizia2020}} &
\textbf{~\cite{Radford2015}} &
\textbf{~\cite{Chien2021}} &
\textbf{NGGAN} \\ \hline  
    \textbf{0-15 kHz} & 12$\%$ & 0$\%$ & 3$\%$ & \textbf{5$\%$} & 3$\%$ \\ \hline
    \textbf{15-35 kHz} & 81$\%$ & 86$\%$ & \textbf{81$\%$} & 53$\%$ & 84$\%$ \\ \hline
    \textbf{35-60 kHz} & 1$\%$ & \textbf{1$\%$} & \textbf{1$\%$} & 3$\%$ & 0$\%$ \\ \hline
    \textbf{60-200 kHz} & 1$\%$ & 3$\%$ & \textbf{2$\%$} & 6$\%$ & \textbf{2$\%$} \\ \hline
    \textbf{Error} & - & 19$\%$ & \textbf{10$\%$} & 42$\%$ & 14$\%$ \\ \hline    
    \end{tabular}}
}    
\label{tab:cyclic_statistic_3}
\end{table}

\begin{figure}[tbh]
  \centering
  \includegraphics[width=.45\textwidth]{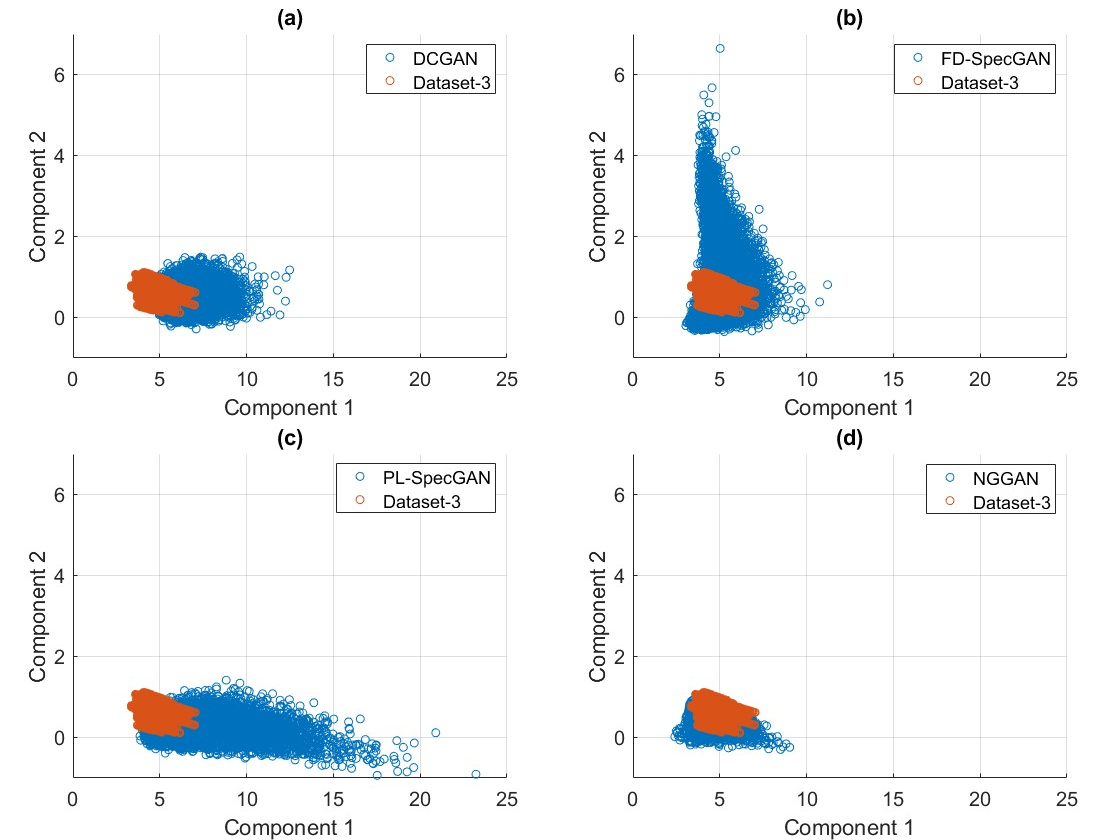}
  \caption{PCA scatter: (a) DCGAN~\cite{Letizia2020}; (b) FD-SpecGAN~\cite{Radford2015}; (c) PL-SpecGAN~\cite{Chien2021}; (d) NGGAN.}
  \label{fig:pca_3}
\end{figure}

Fig.~\ref{fig:pca_3} presents PCA scatter plots of noise generated by each GAN-based model and Dataset-3.
The PCA scatter plots linked to FD-SpecGAN and PL-SpecGAN show heavier disparities than those associated with NGGAN. 
This can result in a significant deviation of the generated data from the distribution of the real data, thereby posing adverse consequences. 
This emphasizes the importance of balancing fidelity and diversity in GAN models.
The noise characteristics generated by the NGGAN were closest to those of Dataset 3. A comparison of the FID values in Table~\ref{tab:FID_3} revealed that the NGGAN provides a suitable balance between the quality and diversity of the generated noise.

\begin{table}[th]
\centering
\caption{PCA feature FID analysis for Dataset-3.}
    \scalebox{1}{
    \begin{tabular}{|c|c|c|c|c|} 
    \hline
    \textbf{DCGAN~\cite{Letizia2020}} &\textbf{FD-SpecGAN~\cite{Radford2015}} &\textbf{PL-SpecGAN~\cite{Chien2021}} &\textbf{NGGAN}
    \\
    \hline
     0.717 & 2.27 & 1.74 & \textbf{0.24}
    \\
    \hline
    \end{tabular}}
\label{tab:FID_3}
\end{table}

The proposed NGGAN consistently demonstrated superior performance compared to the FD-SpecGAN and PL-SpecGAN.
The noise traces produced by the NGGAN closely mirrored the quality and diversity of the measured dataset, establishing it as an effective model for learning and generating noise in NB-PLC systems.
The PCA scatter diagrams confirmed that the samples generated by the NGGAN covered most of the areas obtained by each dataset.
Moreover, the disparities in the results generated from the dataset demonstrate the diversity of the NGGAN.
This suggests that the NGGAN achieves a favorable equilibrium between fidelity and diversity.
The proposed NGGAN can be used to generate noise patterns for evaluating the robustness of NB-PLC receivers against complicated noise in powerline networks~\cite{Rouissi2019}.
Furthermore, the NGGAN is a learnable data augmentation method for training artificial intelligence (AI)-based NB-PLC transceivers.
The concept of learnable data augmentation can be extended to the design of wireless consumer electronic devices that suffer from complex noise~\cite{Lemley2020, Sharma2019}. 

\subsection{Training and Testing Time Complexity Analysis}

In our simulations, Python 3.7 and its associated libraries were utilized to construct the GAN model architecture. We employed an Intel i5-7400 (CPU) and an Nvidia GTX 1080Ti (GPU) as the execution hardware. 
The training dataset consisted of $16,384$ samples with a batch size of 32.
Table~\ref{tab:time_complexity} (a) lists the time required to train each epoch for each GAN-based model using various noise datasets. In general, the DCGAN outperformed the other GAN-based models in terms of training time, but its performance metrics were the worst overall. Notably, the proposed NGGAN demonstrated significant improvements compared to FD-SpecGAN and PL-SpecGAN in terms of training time.
Table~\ref{tab:time_complexity} (b) lists the time required to test the generated data for each GAN-based model using different noise datasets. In general, the trained PL-SpecGAN exhibited the shortest testing time compared to the other GAN-based models. Conversely, the trained DCGAN required the longest testing time among all trained GAN-based models. The proposed NGGAN completes testing in half the time required by the trained DCGAN.

\begin{table}[th]
\centering
\caption{Time complexity analysis: (a) training time per epoch, and (b) testing time per generated data.}
\subfloat[][]{
    \scalebox{0.9}{
    \begin{tabular}{|c|c|c|c|c|} 
    \hline
    \textbf{Dataset} & \textbf{\cite{Letizia2020}} &\textbf{\cite{Radford2015}} & \textbf{\cite{Chien2021}} & \textbf{NGGAN} 
     \\
    \hline
    \textbf{Dataset-1} & \bf{1 min 23 sec} & 12 min 38 sec & 9 min 45 sec & 8 min 14 sec
    \\
    \hline
    \textbf{Dataset-2} & \bf{2 min 29 sec} & 19 min 53 sec & 15 min 41 sec & 7 min 6 sec
    \\
    \hline
    \textbf{Dataset-3} & \bf{1 min 24 sec} & 13 min 58 sec & 15 min 58 sec & 7 min 55 sec 
    \\
    \hline
    \end{tabular}}
}
\\

\subfloat[][]{
    \scalebox{0.95}{
    \begin{tabular}{|c|c|c|c|c|} 
    \hline
    \textbf{Dataset} & \textbf{\cite{Letizia2020}} &\textbf{\cite{Radford2015}} & \textbf{\cite{Chien2021}} & \textbf{NGGAN} 
    \\
    \hline
    \textbf{Dataset-1} & 0.02 sec & 0.005 sec & \bf{0.003 sec} & 0.01 sec
    \\
    \hline
    \textbf{Dataset-2} & 0.02 sec & 0.005 sec & \bf{0.003 sec} & 0.01 sec
    \\
    \hline
    \textbf{Dataset-3} & 0.02 sec & 0.005 sec & \bf{0.003 sec} & 0.01 sec 
    \\
    \hline
    \end{tabular}}
}
\label{tab:time_complexity}
\end{table}

\section{Conclusions}
\label{sec:conclusions}

This study proposes an NGGAN model to learn cyclo-stationary noise in NB-PLC systems using two mathematically modeled noise datasets (Dataset-1 and Dataset-2) and one real measurement dataset (Dataset-3).
We simplified the architecture of the DCGAN by transforming noise data into spectrograms and extending the length of the input data based on the cyclo-stationary property.
The Wasserstein distance was used as the loss function of the NGGAN to enhance the similarity between the generated noise and the three datasets.
Cyclic spectrum and diversity analyses were performed to ensure that the generated noise ensured inference of the data distribution of the real environment from an insufficient practical noise dataset.
In our simulation, the proposed NGGAN consistently outperformed DCGAN, FD-SpecGAN, and PL-SpecGAN.
Specifically, the resulting PCA scatter and FID analysis showed that NGGAN can generate higher fidelity and diversity noise samples than the comparative methods.
Therefore, the proposed NGGAN serves as a data augmentation approach to provide a generated dataset for designing denoising and robustness for NB-PLC transceivers.

\begin{IEEEbiography}[{\includegraphics[width=1in,height=1.25in,clip,keepaspectratio]{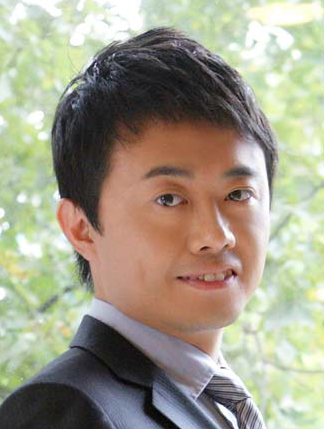}}]{Ying-Ren Chien} (Senior Member, IEEE) received the B.S. degree in electronic engineering from the National Yunlin University of Science and Technology, Douliu, Taiwan, in 1999, and the M.S. degree in electrical engineering and the Ph.D. degree in communication engineering from National Taiwan University, Taipei, Taiwan, in 2001 and 2009, respectively. 

Dr. Chien joined the Department of Electrical Engineering, National Ilan University (NIU), Yilan City, Taiwan from 2012 to 2025. He has been promoted to Full Professor since 2018; he served as the Chair at NIU from 2018 to 2025. Since 2025, he has been with the Department of Electronic Engineering, National Taipei University of Technology (NTUT), Taipei, Taiwan, where he is currently a Full Professor. From 2023 to 2024, he was the vice chair of IEEE Consumer Technology Society (CTSoc) Virtual Reality, Augmented Reality, and Metaverse (VAM) Technical Committee (TC). Since 2025, he has been the Secretary of IEEE CTSoc Audio/Video Sytems and Signal Processing (AVS) TC.

Dr. Chien is currently an Associate Editor for the IEEE TRANSACTIONS ON CONSUMER ELECTRONICS.
He received Best Paper Awards, including ICCCAS 2007, ROCKLING 2017, and IEEE ISPACS 2021. Dr. Chien was presented with the IEEE CESoc/CTSoc Service Awards (2019), NSC/MOST Special Outstanding Talent Award (2021, 2023, 2024), Excellent Research-teacher Award (2018 and 2022), and Excellent Teaching Award (2021). His research interests are consumer electronics, multimedia denoising algorithms, adaptive signal processing theory, active noise control, machine learning, Internet of Things, and interference cancellation.
\end{IEEEbiography}

\begin{IEEEbiography}[{\includegraphics[width=1in,height=1.25in,clip,keepaspectratio]{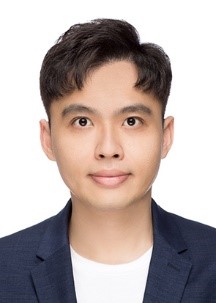}}]{Po-Heng Chou}(Member, IEEE) was born in Tainan, Taiwan. He received the B.S. degree in electronic engineering from National Formosa University (NFU), Huwei, Yunlin, Taiwan, in 2009, the M.S. degree in communications engineering from National Sun Yatsen University (NSYSU), Kaohsiung, Taiwan, in 2011, and the Ph.D. degree from the Graduate Institute of Communication Engineering (GICE), National Taiwan University (NTU), Taipei, Taiwan, in 2020. His research interests include AI for communications, deep learning-based signal processing, wireless networks, and wireless communications, etc.

He was a Postdoctoral Fellow at the Research Center for Information Technology Innovation (CITI), Academia Sinica, Taipei, Taiwan, from Sept. 2020 to Sept. 2024. 
He was a Postdoctoral Fellow at the Department of Electronics and Electrical Engineering, National Yang Ming Chiao Tung University (NYCU), Hsinchu, Taiwan, from Oct. to Dec. 2024.
He has been elected as the Distinguished Postdoctoral Scholar of CITI by Academia Sinica from Jan. 2022 to Dec. 2023. He is invited to visit Virginia Tech (VT) Research Center (D.C. area), Arlington, VA, USA, as a Visiting Fellow, from Aug. 2023 to Feb. 2024.
He received the Partnership Program for the Connection to the Top Labs in the World (Dragon Gate Program) from the National Science and Technology Council (NSTC) of Taiwan to perform advanced research at VT Institute for Advanced Computing (D.C. area), Alexandria, VA, USA, from Jan. 2025 to present.

Additionally, Dr. Chou received the Outstanding University Youth Award and the Phi Tau Phi Honorary Membership from NTU in 2019 to honor his impressive academic achievement. He received the Ph.D. Scholarships from the Chung Hwa Rotary Educational Foundation from 2019 to 2020.
\end{IEEEbiography}

\begin{IEEEbiography}[{\includegraphics[width=1in,height=1.25in,clip,keepaspectratio]{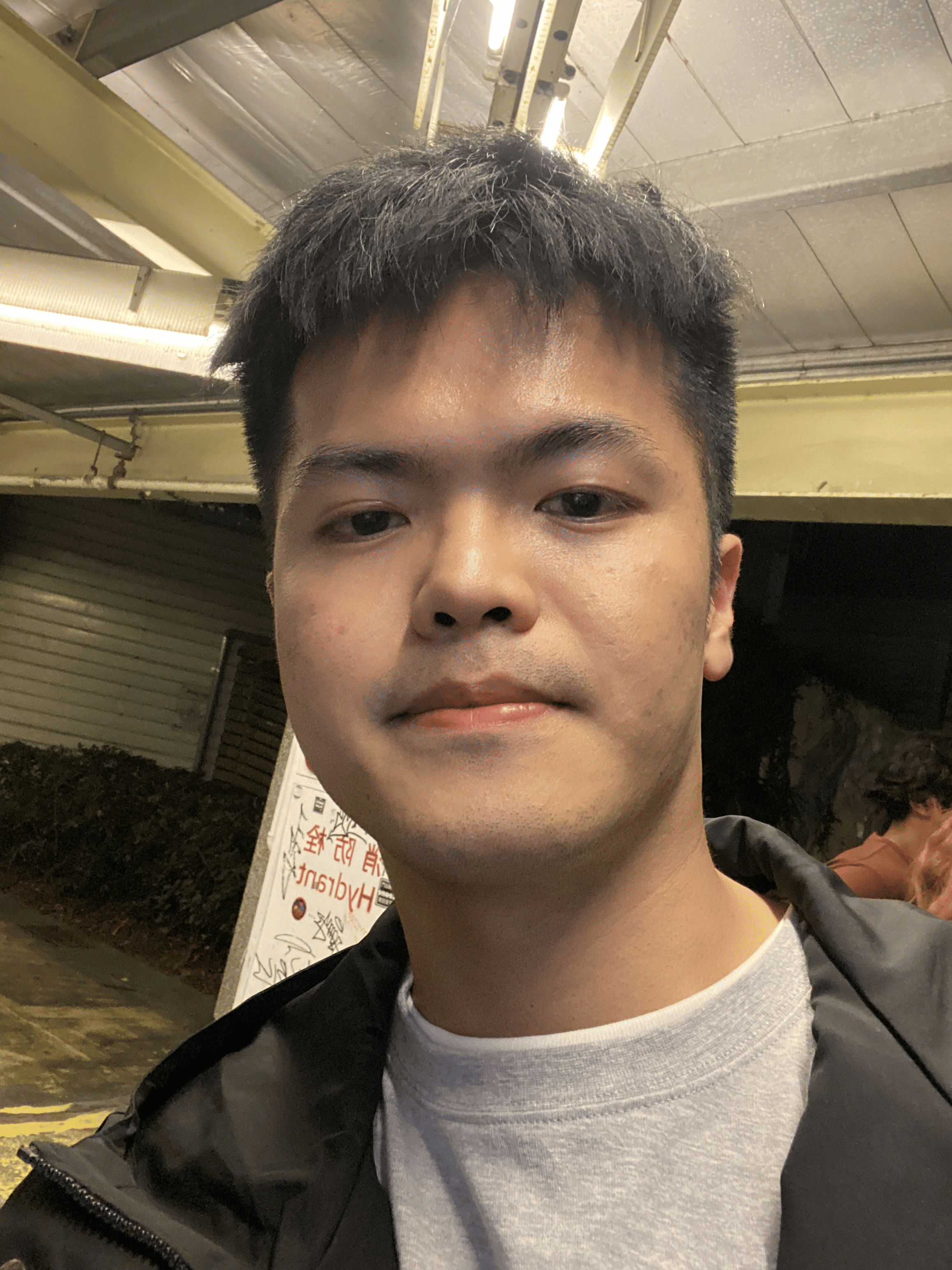}}]{You-Jie Peng} was born in Beipu, Hsinchu City, Taiwan, in 1996.
He received the B.E. degree from National Taipei University (NTPU), Taipei, Taiwan, in 2020 and the M.S. degree from the Graduate Institute of Communication Engineering (GICE), National Taiwan University (NTU), Taipei, Taiwan, in 2022.

His current research interests include deep learning, statistical signal processing, and powerline communications.
He won the Best Paper Award of the International Symposium on Intelligent Signal Processing and Communication Systems in 2021.
\end{IEEEbiography}

\begin{IEEEbiography}[{\includegraphics[width=1in,height=1.25in,clip,keepaspectratio]{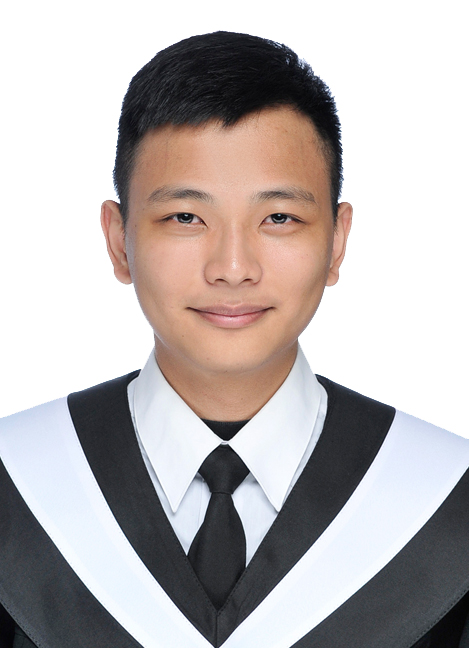}}]{Chun-Yuan Huang} was born in Changhua, Taiwan, in 2000. He received the B.S. degree in computer and communications engineering from the National Kaohsiung University of Science and Technology (NKUST), Kaohsiung, in 2022, and the M.S. degree in communications engineering from National Sun Yat-sen University (NSYSU), Kaohsiung, in 2024. His research interests include deep learning for wireless communications and communication theory.
\end{IEEEbiography}

\begin{IEEEbiography}[{\includegraphics[width=1in,height=1.25in,clip,keepaspectratio]{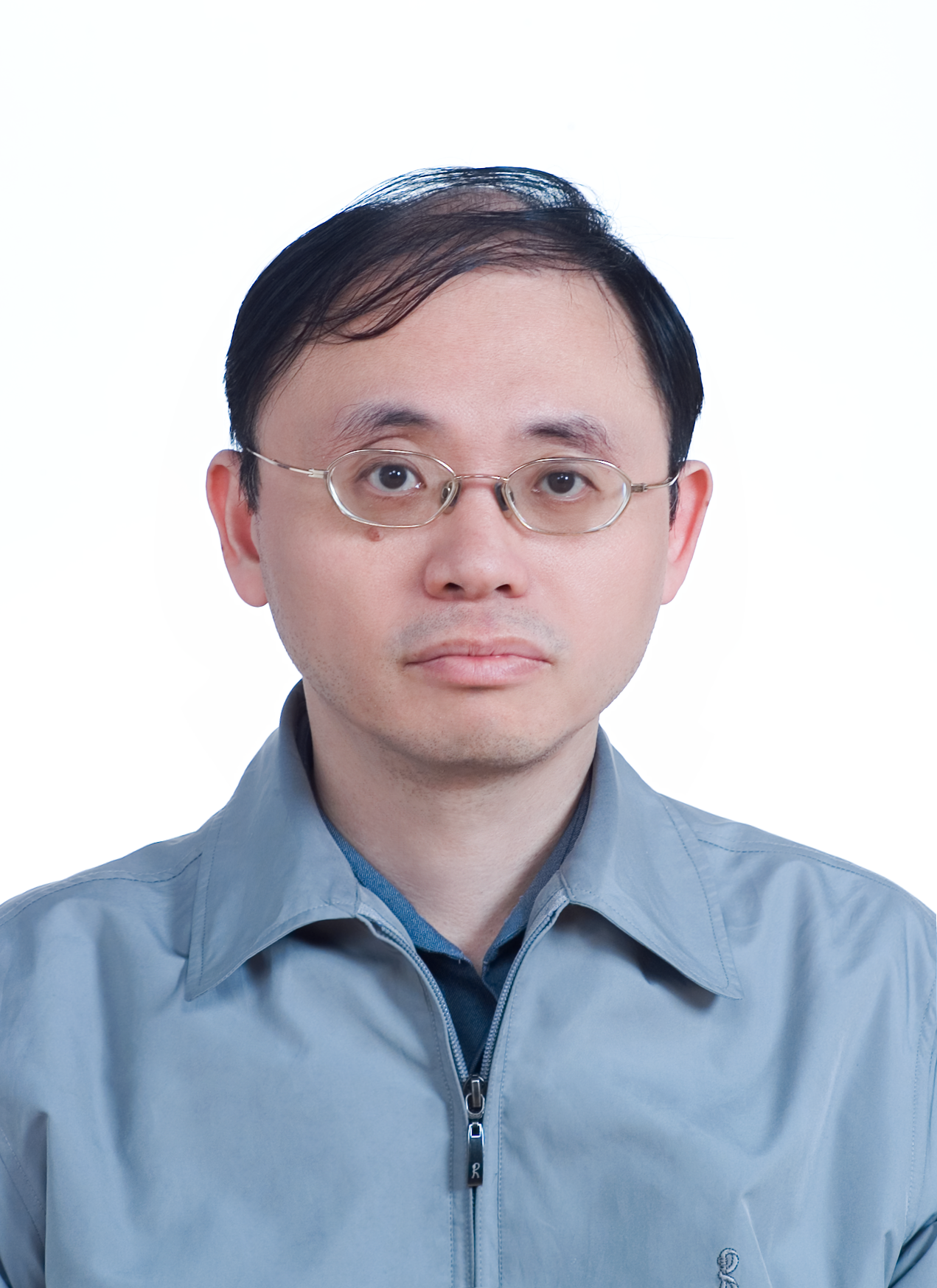}}]{Hen-Wai Tsao} was born in Taipei, Taiwan, in 1953. He received the B.S., M.S., and Ph.D. degrees from National Taiwan University (NTU), Taipei, Taiwan, in 1975, 1978, and 1990, respectively, all in electrical engineering. 

Since 1978, he has been with the Department of Electrical Engineering, National Taiwan University, where he is currently a Professor Emeritus. His main research interests include broadband communication systems, communication electronics, instrumentation systems, and related electronic circuits.
\vspace{1.3in}
\end{IEEEbiography}

\begin{IEEEbiography}[{\includegraphics[width=1in,height=1.25in,clip,keepaspectratio]{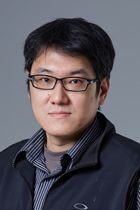}}]{Yu Tsao}(Senior Member, IEEE) received his B.S. and M.S. degrees in Electrical Engineering from National Taiwan University, Taipei, Taiwan, in 1999 and 2001, respectively, and his Ph.D. degree in Electrical and Computer Engineering from the Georgia Institute of Technology, Atlanta, GA, USA, in 2008. 

From 2009 to 2011, he was a researcher at the National Institute of Information and Communications Technology, Kyoto, Japan, where he worked on research and product development in automatic speech recognition for multilingual speech-to-speech translation. He is currently a Research Fellow (Professor) and the Deputy Director of the Research Center for Information Technology Innovation at Academia Sinica, Taipei, Taiwan. He also serves as a Jointly Appointed Professor in the Department of Electrical Engineering at Chung Yuan Christian University, Taoyuan, Taiwan. His research interests include assistive oral communication technologies, audio coding, and bio-signal processing. 

Dr. Tsao is currently an Associate Editor for the IEEE/ACM TRANSACTIONS ON AUDIO, SPEECH, AND LANGUAGE PROCESSING and IEEE SIGNAL PROCESSING LETTERS. He was the recipient of the Academia Sinica Career Development Award in 2017, National Innovation Awards from 2018 to 2021, the Future Tech Breakthrough Award in 2019, the Outstanding Elite Award from the Chung Hwa Rotary Educational Foundation in 2019–2020, the NSTC FutureTech Award in 2022, and the NSTC Outstanding Research Award in 2023. He is the corresponding author of a paper that received the 2021 IEEE Signal Processing Society (SPS) Young Author Best Paper Award.
\end{IEEEbiography}

\end{document}